\documentclass[usenatbib, useAMS]{mn2e}
\input epsf

\newcommand{\lapprox}{\lower0.8ex\hbox{$\buildrel <\over\sim$}}
\newcommand{\gapprox}{\lower0.8ex\hbox{$\buildrel >\over\sim$}}
\usepackage{epsfig}
\usepackage{epsf}
\usepackage{times}
\input epsf
\def \ul #1:#2:{$^{+#1}_{-#2}$}

\title[Mocassin]{Mocassin: A fully three-dimensional Monte Carlo photoionization code}
\author[Ercolano et al. ]
{B. Ercolano$^1$, M. J. Barlow$^1$, P. J. Storey$^1$, X.-W. Liu$^{1,2}$\\
$^1$University College London, Gower Street, London WC1E 6BT, UK\\
$^2$Current address: Department of Astronomy, Peking University, Beijing 100871, P. R. China
}
\date{Received:}
\begin{document}
\maketitle
\begin{abstract}
\noindent

The study of photoionized environments is fundamental to many astrophysical problems. Up to the present most photoionization codes have numerically solved the equations of radiative transfer by making the extreme simplifying assumption of spherical symmetry. Unfortunately very few real astronomical nebulae satisfy this requirement. To remedy these shortcomings, a self-consistent, three-dimensional radiative transfer code has been developed using Monte Carlo techniques. The code, Mocassin, is designed to build realistic models of photoionized nebulae having arbitrary geometry and density distributions, with both the stellar and diffuse radiation fields treated self-consistently. In addition, the code is capable of treating ones or more exciting stars located at non-central locations.

The gaseous region is approximated by a cuboidal Cartesian grid composed of numerous cells. The physical conditions within each grid cell are determined by solving the thermal equilibrium and ionization balance equations. This requires a knowledge of the local primary and secondary radiation fields, which are calculated self-consistently by locally simulating the individual processes of ionization and recombination. The structure and the computational methods used in the Mocassin code are described in this paper.

Mocassin has been benchmarked against established one-dimensional spherically symmetric codes for a number of standard cases, as defined by the Lexington/Meudon photoionization workshops \citep{pequignot86,ferland95, pequignot01}. The results obtained for the benchmark cases are satisfactory and are presented in this paper. A performance analysis has also been carried out and is discussed here.

\end{abstract}

\begin{keywords}
H~{\sc ii} regions -- planetary nebulae: general -- ISM: abundances -- atomic processes

\end{keywords}

\section{Introduction}
Amongst the first numerical models for photoionized gaseous nebulae were those calculated by \citet{flower68}, \citet{harrington68} and \citet{rubin68}. These early models included the basic physical 
processes of ionization and recombination of hydrogen and helium, thermal balance and escape of the emitted photons from the nebula. However, the lack of reliable atomic data 
heavily limited the success of these models, as well as the fact that a number of important physical processes, such as charge exchange and dielectronic recombination \citep{aldrovandi73, pequignot78, storey81}, were not accounted for at the time. The evolution of photoionization modelling has gone hand in hand with advances made in atomic 
physics and computer technology. The application of photoionization models to a wider range of ions has been aided by the photoionization cross-section calculations by \citet{reilman79}, and, more recently, the Opacity Project \citep{hummer93}. Compilations based on the latter's data \citep[e.g.][]{verner95}, have made possible the inclusion of accurate photoionization cross-sections for many more ions in calculations. \citet{mendoza82} presented a compilation of radiative and 
collisional data for collisionally excited ultraviolet, optical and infrared lines which was widely adopted, with some of these data still in use today, though most have been seperceded by more recent calculations such as the R-matrix calculations of the Iron Project \citep{hummer93} and the Belfast group \citep[e.g.][]{mclaughlin98, ramsbottom98}.  Currently, radiative and dielectronic recombination rates are still highly uncertain or unavailable for some ions; recent efforts to improve the situation have been reviewed by \citet{nahar99} and \citet{nahar00}. Most photoionization models include temperature-dependent analytical fits to these recombination rates, such as those of \citet{aldrovandi73} for radiative and high temperature dielectronic recombination, and those of \citet{nussbaumer83} for low temperature dielectronic recombination. 

Available computer power has increased enormously since the dawn of photoionization modelling. This has allowed more complex models to be built, including more ions,  more frequency points, more lines and more atomic levels. Nevertheless, the fundamental assumption of spherical symmetry has always been retained. However, a glance at an image of any Galactic H~{\sc ii} region will immediately demonstrate that these objects are neither spherically symmetric nor homogeneous. In addition, they usually have multiple exciting stars located at non-central positions in the nebula. By contrast, planetary nebulae (PNe) have only a single, centrally located, exciting star. However, even for PNe, spherical symmetry is not a realistic assumption, as demonstrated by observations with instruments such as the Hubble Space Telescope, which reveal an overwhelming variety in the 
shapes of planetary nebulae. These objects are very rarely circular in projection; a recent study inferred that about 50\% of all known planetary nebulae are low eccentricity ellipticals, while only about 10\% are circular in projection, with the remainder having more extreme
elliptical or bipolar geometries \citep{soker97, soker01}. Some objects, for example the two young planetary nebulae He~2-47 and PN~M1-37, \citep[also dubbed the {\it starfish twins;}][]{sahai00}, show even more complicated geometries, with multiple  lobes. Other PNe have FLIERs \citep[fast, low ionization emitting regions;][]{balick93, balick94, balick98}, BRETS \citep[bipolar, rotating, episodic jets; e.g.][]{lopez93}, ansae, jets, knots, filaments, tails or multiple envelopes. \citep[see e.g.][]{perinotto00, corradi99, garcia97}. 

To our knowledge, only two three-dimensional photoionization codes have been develped so far, one by \citet{bassgen90} and the other by Gruenwald, Viegas \& de Broguiere (1997). The first code used a fixed number of equally sized cells and the on-the-spot approximation for the diffuse radiation field, with only the six more abundant chemical elements being taken into account. The work by \citet{gruenwald97} improves on this by allowing a more flexible spatial grid and by using an iterative technique for the determination of the diffuse field and also by including twelve chemical elements in the simulations. 

Since most existing one-dimensional photoionization codes are  based on the numerical solution of the equations of radiative transfer assuming spherical symmetry, their expansion to three dimensions can be either very difficult or impractical, resulting in very large codes. The Monte Carlo approach to transfer problems provides a geometry-independent technique which can handle the radiation transport problem self-consistently. With this in mind, the Mocassin code (MOnte CArlo SimulationS of Ionised Nebulae) was developed, in order to provide a three-dimensional modelling tool capable of dealing with asymmetric and/or inhomogeneous nebulae, as well as, if required, multiple, non-centrally located exciting stars.  

Section~\ref{sec:description} contains a description of the general Mocassin architecture and of some of the main computational methods used in the code. The code has been benchmarked against established spherically symmetric one-dimensional photoionization codes for a set of standard nebulae and in Section~\ref{sec:benchmarks} we present the results of this benchmarking, together with a performance analysis of the codes. In section~\ref{sec:discussion} we discuss the results of the benchmarking and present some general guidelines on how to run the code efficiently. 

\section{Description of the Monte Carlo code}
\label{sec:description}

\subsection{Background}
The Monte Carlo method has been widely applied to a variety of astrophysical problems, such as the penetration of ultraviolet radiation into the interiors of uniform or lumpy, interstellar clouds \citep{flannery80, boisse90}, resonance-like scattering in accretion disc winds \citep{knigge95} and polarization maps for the circumstellar envelopes of protostars \citep{fischer94}. In the examples described above the absorption and scattering coefficients are not coupled to the radiation field and, therefore, these problems do not require solution by iteration. 

However, Monte Carlo techniques have also been used for dust radiative equilibrium calculations for some time, see e.g. \citet{lefevre82}, \citet{lefevre83} and, more recently, \citet{wolf99}. These authors use a technique in which stellar and diffuse photon packets are emitted separately; the number of diffuse photon packets (i.e. packets emitted by the dust) is determined by the dust grain temperature, which in turn is determined by the balance between the number of absorbed and emitted photon packets. An initial guess for the dust grain temperature is provided by the number of packets absorbed, and the iteration continues until the 
grain temperatures converge. Using this method the stellar luminosity is not automatically conserved during the Monte Carlo simulation; only after the grain temperatures have reached convergence is the stellar luminosity approximately conserved. The convergence of such codes is often very slow and requires a large number of iterations and simulation quanta in order to reach the required accuracy. 
 
\citet{bjorkman01} have described a general 
radiative equilibrium and temperature correction procedure 
for use in  Monte Carlo radiative transfer codes having sources of temperature-independent opacity, such as dust. Their technique makes use of information 
naturally given by  the Monte Carlo method, which, by tracking every photon/energy packet, makes it easy
to determine where in the simulation grid energy is being absorbed.  When energy is deposited at a given 
location, following a packet's absorption, the local medium is heated.  Whenever this occurs the new local 
temperataure is calculated and the packet is then re-emitted accordingly.  The packets are followed in
their path through the region, as they undergo scatterings and absorptions followed by re-emissions, with the temperatures being updated after each event,
 until the packets reach the edge of the nebula and escape to infinity, hence contributing to the emergent spectrum.  Once all the stellar photon packets have escaped, the resulting envelope temperature and the emergent spectrum are correct without the need of any further iterations.

A great limitation of Bjorkman \& Wood's method is that it cannot be applied to situations where the 
opacities are temperature-dependent, as is the case in photoionized nebulae.
 There are two reasons for the failure of this method when the opacity varies with the local temperature:
firstly,  the number of photon packets absorbed by the cell prior to a temperature update would 
be either too small or too large, and, secondly, a change in temperature would also imply a change of the interaction locations of previous packets, signifying that the paths of the previous photon packets should have been different.  While, it is clear that, when dealing with photoionised gas,   Bjorkman \& Wood's technique
is not applicable, their work is nevertheless very enlightening and should be taken into account for 
further developments of the Mocassin code, when a treatment for dust grains will be introduced.  

A recent example of the application of the Monte Carlo technique to problems requiring solution by iteration is the work of \citet{lucy99}, who obtained the temperature stratification and emergent spectrum of a non-grey spherically symmetric extended stellar atmosphere in LTE. His results show very good agreement with the predictions of \citet{castor74}, hence demonstrating the validity of the Monte Carlo techniques applied, some of which were also used in the development of Mocassin. 
The current work folows the approach described by \citet{lucy99} and also applied in the one-dimensional code developed by Och, Lucy \& Rosa (1998). They employed a different Monte Carlo treatment of the radiative transfer in order to iteratively determine the temperature and ionisation stratification  for a 
spherically symmetric photoionised nebula of uniform density. Some of the techniques that they used are also described in detail by \citet{lucy99, lucy01, lucy02}. The basic concept is that, when calculating radiative equilibrium
temperatures, conservation of stellar luminosity is more important than the details of the spectral energy distribution.
With this in mind conservation of stellar luminosity is enforced by using energy packets of constant net energy throughout the simulations.
Moreover, all absorbed packets are re-emitted immediately after every absorption event.  The 
frequencies of the re-emitted energy packets are determined by the local gas emissivities.  Although the frequency 
distribution of the re-emitted packets will not be correct until the nebular temperatures have converged, 
this method naturally enforces radiative equlibrium at each point in the nebula and so naturally provides
conservation of energy.  This not only results in a simpler code but also makes the convergence of the gas temperatures easier \citep{lucy99, lucy01}.  
Energy packets will be discussed in more detail in section ~\ref{sub:energyPackets}.

\subsection{Energy Packets}
\label{sub:energyPackets}
The main principle of our treatment of a photoionized nebula consists of locally simulating the individual processes of ionization and recombination.  The radiation field is
therefore expressed in terms of energy packets, $\varepsilon(\nu)$, which are the calculation quanta.  $\varepsilon(\nu)$ is a packet consisting of $n$ photons of frequency 
$\nu$ such that 
\begin{equation}
\varepsilon (\nu) = nh\nu
\end{equation}
In addition, we take all packets to have constant energy $\varepsilon_0$.  There are several reasons for choosing to work with monochromatic, indivisible packets of 
radiant energy instead of photons.  First of all, energy packets are more computationally economic and, also, since they all have the same energy, 
then those packets emitted in the infrared will contain a larger number of photons which, as a consequence, will not have to be followed individually \citep{abbott85}. Note that all energy packets are followed until they escape the nebula, including infrared energy packets. This is in order to allow the introduction of dust particles into the radiative transfer treatment of Mocassin, which is planned for the near future.
Also, as the total stellar luminosity, $L_*$, is evenly split amongst the stellar energy packets, the energy carried by a single packet in the time interval $\Delta \, t$, 
which represents the duration of the Monte Carlo experiment, is given by
\begin{equation}
\frac{L_*}{N} = \frac{\varepsilon_0}{\Delta t}
\label{eq:enpacket}
\end{equation}
where $N$ is the number of energy packets used in the simulation \citep{och98}.
Most importantly, the use of constant energy packets is a natural way of imposing strict energy conservation at any point in the nebula \citep{lucy99}.  
So, when a packet of radiant energy $\varepsilon(\nu_a)~=~\varepsilon_0$ is absorbed, it is immediately re-emitted with  a frequency $\nu_{\rm e}$,  which is determined according to a 
frequency distribution set by the gas emissivity of the current cell. The packet emitted,  $\varepsilon(\nu_{\rm e})$, will then have the same energy as the absorbed packet, 
$\varepsilon(\nu_a)$, meaning that only the number, $n$, of photons contained in the packet is changed. 

\subsection{Initiation}
\label{sub:initiation}
In our modelling the gaseous region is approximated by a three-dimensional Cartesian grid, where the ionising source can be placed at the centre of the grid or anywhere else in the grid. 
This feature is very useful when dealing with axisymmetric nebulae, since, by placing the source in a corner of the grid, we need only consider one eighth of the 
nebula, which can then be reconstructed in full at the end of the simulation. This allows the running of models with much higher spatial resolution than those which would be possible if 
a full nebula had to be considered, by putting the source in the centre and, therefore, not making use of any symmetry properties of the object. Switches built inside the code 
allow the user to 
specify whether the nebula has some degree of symmetry and, if so, whether the symmetry is to be used. 

Inside each grid cell all nebular properties, such as the mass density of the gas, $\rho$; the electron temperature and density, $T_{\rm e}$ and $N_{\rm e}$; and the frequency dependent gas opacity and 
emissivity, $\kappa_{\nu}$ and $j_{\nu}$, are constant by definition. Thermal balance and ionisation equlibrium are imposed in each grid cell in order to obtain 
the physical conditions in the local gas. 

The energy packets are created at the position of the ionising source and they all carry the same energy $\varepsilon_0$,  as discussed in the previous section. 
The frequency, $\nu$, of each individual packet emitted is derived from the input spectrum of the ionising source according to the probability density function 
\begin{equation}
p(\nu) = \frac{F_{\nu}d\nu}{\int_{\nu_{min}}^{\nu_{max}}F_{\nu'}d\nu '} = \frac{F_{\nu}d\nu}{L_*/(4\pi R^{2}_*)}
\end{equation}
where $F_{\nu}$ is the stellar flux and $R_*$ is the stellar radius. This is then the probability of an energy packet being emitted with a frequency lying in the interval 
$(\nu, \nu+d\nu)$. The upper and lower integration limits, $\nu_{min}$ and $\nu_{max}$ , have to be chosen properly, depending on the input spectrum, in order to
ensure that the bulk of the radiation is included in the frequency range. As the source emits energy isotropically, the direction of travel of every energy packet emitted is chosen randomly. This is done by choosing two random numbers, $\alpha$ and $\beta$, in the interval $[0,1]$, and calculating the following quantities:
\begin{eqnarray}
w      & = & 2\alpha - 1        \nonumber \\
t      & = & \sqrt{1-w^2}       \nonumber \\
\theta & = & \pi (2\beta - 1)   \nonumber \\
u      & = & t cos \theta       \nonumber \\
v      & = & t sin \theta     
\label{eq:randomDirection}
\end{eqnarray}
The random unit vector in Cartesian coordinates is then $(u,v,w)$ \citep{harris97}.

\subsection{Trajectories}
\label{sub:trajectories}
Once a stellar packet is created at the source and launched into the nebula, its trajectory must be computed as it undergoes absorptions followed by re-emissions due 
to bound-free and free-free processes. The trajectory ends when the packet reaches the edge of the nebula, where it escapes to infinity and contributes to the 
emergent spectrum. 

There are two methods to track the packets and determine the locations of the absorption events. Consider a 
packet of frequency $\nu_{p}$,  emitted in the direction $\hat{u}$. 
The first of these methods consists of calculating the run of optical depth, $\tau_{\nu_{\rm p}}$, at the energy packets' frequency $\nu_{\rm p}$,
from the location at which the packet is emitted to the edge of the ionised region along the direction of travel, $\hat{u}$. The probability of absorption along that path is then given by 
\begin{equation}
p(\tau_{\nu_{\rm p}}) = {\rm e}^{-\tau_{\nu_{\rm p}}}
\label{eq:p}
\end{equation}
and the normalised cumulative probability function is given by
\begin{eqnarray}
P(l) & = &\frac{\int_{0}^{\tau_{\nu_{\rm p}}(l)}e^{-\tau_{\nu_{\rm p}}}d\tau_{\nu_{\rm p}}}{\int_{0}^{\infty}{\rm e}^{-\tau_{\nu_{\rm p}}}d\tau_{\nu_{\rm p}}} \nonumber \\
      & = &1 - {\rm e}^{-\tau_{\nu_{\rm p}}(l)}
\end{eqnarray}
where $\tau_{\nu_{\rm p}}(l)$ is the optical depth to the absorption event and $l$ is the path length.
The position at which the energy packet will be absorbed will then be determined by choosing a random number in the interval $[0, 1]$ and comparing it against $P(l)$. In reality,
 it is more convenient to use the inverse approach, where the optical depth from the energy packet source to the event can be derived from the inverse of equation~\ref{eq:p}
\begin{equation}
\tau_{\nu_{\rm p}}(l) = -{\rm ln}(1 - U_R)
\label{eq:tau}
\end{equation}
where $U_R$ is a random number in the interval $[0, 1]$. Once $\tau_{\nu_{\rm p}}(l)$ has been calculated then the path length can be directly derived \citep{harris97}.

The second method was suggested by \citet{lucy99} and it consists of testing whether an absorption event occurs, on a cell by cell basis. 
In other words, assume that, within each uniform cell, the random path of a packet between events is 
given by equation~\ref{eq:tau}, which corresponds to a physical displacement, $l$, given by
\begin{equation}
\tau_{\nu_{\rm p}} = \kappa_{\nu}\rho \, l
\label{eq:taucell}
\end{equation}
where $\kappa_{\nu}$ and $\rho$ are the frequency dependent absorption coefficients and the density of the current cell 
respectively. The method then
consists of checking whether the displacement $l$ is large enough to carry the packet out of its current cell. 
If this is the case, the packet is moved along its direction of travel, $\hat{u}$, up 
to the boundary with the adjacent cell, where a new value for $U_R$ is cast, giving a new $\tau_{\nu_{\rm p}}$,
 and any further movement of the packet in this new cell is to be followed. Alternatively, if the 
displacement $l$ is not large enough to carry the energy packet across the next boundary, the packet will be 
absorbed and then re-emitted at the end-point of the displacement. Lucy also clarifies in his paper that 
the selection of a new value of $\tau_{\nu_{\rm p}}$ at the crossing of a boundary does not introduce a bias since a photon 
always has an expected path length to its next event corresponding to $\tau_{\nu}=1$, regardless of the 
distance it might already have travelled.

In this work both methods were implemented in the code, in turn, in order to test their respective 
performances. 
The first method proved to be much more computationally expensive then the second. This is due to the fact 
that,  in order to track down the position at which an energy packet is absorbed, using our knowledge of 
$\tau_{\nu_{\rm p}}(l)$, an array 
searching routine has to be used to locate the index of $\tau_{\nu_{\rm p}}(l)$ within the array of optical 
depths calculated from the packet's source to the edge of the nebula. Although the searching procedure employs a 
bisection technique, which makes it quite efficient, the large number of calls to it, due to the large number of energy packet interactions 
within a simulation, means that nearly 60\% of the run time is spent carrying out these searches. 
The second method does not require any calls to the array searching routine, as the packets are followed 
step by step through the nebula, and this results in  the run time being considerably reduced. The current version of Mocassin therefore uses Lucy's appoach to track the energy packets throughout the nebula. 

Finally, the direction of travel of the newly emitted diffuse energy packets (i.e. those packets re-emitted immediately after an absorption event) needs to be determined. Since absorption and re-emission are two independent events, the diffuse packets are emitted isotropically and therefore their direction of travel is chosen randomly using equations \ref{eq:randomDirection}

\subsection{The Mean Intensity}
\label{sub:meanIntensity}
The success of a Monte Carlo model often relies on the careful choice of appropriate {\it estimators}. Monte Carlo estimators provide the means to relate the quantities we 
{\it observe} during our Monte Carlo experiment to the physical quantities we want to determine. In a photoionization model, a measure of the radiation field is needed, namely the mean intensity, $J_{\nu}$. 

In the work of \citet{och98}, the Monte Carlo estimator of $J_{\nu}$ is constructed by using the definition of the specific intensity, $I_{\nu}$, in spherical coordinates, ($r$, $\theta$), as a starting point:
\begin{equation}
\Delta \, E = I_{\nu}(r, \theta) \Delta \, A \mid {\rm cos}\theta \mid \Delta \, \nu \Delta \, \omega \Delta \, t
 \label{eq:ioch}
\end{equation}
where $\Delta \, A $ is the reference surface element, $\theta$ is the angle between the direction of light propagation and the normal to the surface $\Delta\,A$ and $\Delta \, \omega$ is the solid angle.
The mean intensity can then be obtained from this by calculating the zero order moment of $I_{\nu}$, which  gives
\begin{equation}
4 \pi J_{\nu}(r)  =  \int_{\Omega}I_{\nu}\, d \, \omega  =  \frac{\Delta \, E}{\Delta \, t} \sum_{i=1}^{N_{\rm k}}\frac{1}{{\rm cos}\theta_{\rm i}} \frac{1}{\Delta \, A} \frac{1}{\Delta \, \nu}
\label{eq:joch}
\end{equation}
by comparison with equation~\ref{eq:ioch}. The sum is over all packets $N_{\rm k}$ with frequency lying in the interval $(\nu, \nu +d \, \nu)$, crossing $\Delta \, A$ at an angle $\theta$. As discussed above, $\Delta \, E / \Delta \, t$ represents the energy carried by a single packet in the time interval $\Delta \, t$, since  $\Delta\,E~=~\varepsilon_0$, which is given by equation \ref{eq:enpacket}. Equation~\ref{eq:joch} then provides a relation between the Monte Carlo {\it observables} (i.e. the number of energy packets with frequency lying in the interval $(\nu, \nu +d \, \nu)$,  crossing $\Delta \, A$ at angle $\theta$ and the mean intensity of the radiation field, which is the required physical quantity. 

The use of Och et al.'s estimators for $J_{\nu}$, however, becomes problematic in the non-spherically symmetric case, since  the reference surface for the volume elements in an arbitrary two- or three-dimensional coordinate system might not be unique or as obvious as in the one-dimensional case. In our work, a more general expression for the estimator of $J_{\nu}$ is sought, and, therefore, following Lucy's argument \citep{lucy99}, an estimator for $J_{\nu}$ is constructed starting from the result that the energy density of the radiation field in the frequency interval $(\nu, \nu + d\, \nu)$ is $4 \pi J_{\nu} d\nu / c$. At any given time, a packet contributes energy $\varepsilon(\nu) = \varepsilon_0$ to the volume element which contains it. Let $l$ be a packet's path length between successive {\it events}, where the crossing of cell boundaries is also considered an event; the contribution to the time averaged energy content of a volume element, due to the $l$ fragments of trajectory, is $\varepsilon _0 \delta \, t / \Delta \, t$, where $\delta \, t = l/c$. From this argument it follows that the estimator for the volume element's energy density can be written as 
\begin{equation}
\frac{4 \pi J_{\nu} d\nu}{c} = \frac{\varepsilon _0}{\Delta \, t} \frac{1}{V} \sum_{d \, \nu} \frac{l}{c}
\label{eq:jlucy}
\end{equation}
where $V$ is the volume of the current grid cell and the summation is over all the fragments of trajectory, $l$, in $V$, for packets with frequencies lying in the interval $(\nu, \nu + d\,\nu)$. Again, a relation between Monte Carlo observables (i.e. the flight segments, $l$) and the mean intensity of the radiation field, $J_{\nu}$ has been obtained. Moreover, equation~\ref{eq:jlucy} is completely independent of the coordinate system used and, indeed, of the shapes of the volume elements, $V$.  Another important aspect of this approach is that all packets passing through a given grid cell contribute to the local radiation field even without being absorbed; this means that equation~\ref{eq:jlucy} returns estimators of the radiation field even in  the extremely optically thin case when all packets pass through the nebula without any absorption events. From this argument it follows that this technique allows a much better sampling and, hence, in general, much less noisy results compared to other techniques based on estimators for which only packets {\it absorbed} within a given volume element count.

\subsection{Gas emissivity and the diffusion of energy packets}
As we have already discussed in previous sections, after an energy packet is absorbed, a new packet is re-emitted from the same location in a random direction. 
The frequency of the re-emitted packet is calculated by sampling the
spectral distribution of the total local emissivity, $j_{\nu}^{tot}$.  In order to satisfy
the thermal balance implied by the Monte Carlo model, all major emission processes have to be
taken into account, including the complete non-ionizing nebular continuum and line emission, since
 they are part of the energy budget.  The non-ionizing radiation generated in the nebula is assumed to escape without further interaction and  constitutes the {\it observable spectrum} which can then
be compared with observations.  The
following paragraphs are concerned with the description of the individual contributions to the total
emissivity.

The continuum emission due to H~{\sc i}, He~{\sc i}, He~{\sc ii} and to heavier ions is
included. The H~{\sc i} continuum
can be divided into the Lyman continuum, which is capable of ionizing H, and the Balmer, Paschen,
etc.\, continua, which are not capable of ionizing H.  The emissivity in the Lyman continuum is
calculated directly from a combination of the Saha and Milne relations:
\begin{equation}
j_{\nu} =
\frac{h\nu^3}{c^2}\frac{\omega_i}{\omega_{i+1}}(\frac{h^2}{2\pi mkT_{\rm e}})^{3/2}a_{\nu}({\rm X}^i)e^{-h(\nu-\nu_0)/kT_{\rm e}}{\rm X}^{i+1}N_{\rm e}
\label{eq:milne}
\end{equation}
where $\omega_i$ and $\omega_{i+1}$ are the ground state statistical weight of the ions
involved, X$^{i+1}$ is the abundance of the recombining ion, $a_{\nu}({\rm X}^i)$ is the photoionization
cross section and $\nu_0$ is the photoionization threshold. 
The emissivity of the other series continua are obtained by interpolation of published data \citep{ferland80}.  
A similar approach is used for the He~{\sc i} and the He~{\sc ii} continua, where for
frequencies greater
than 1.8 Ryd and 4.0 Ryd, respectively, equation~\ref{eq:milne} is used,
and
the
emissivities at lower frequencies are obtained by interpolation of the data published by
\citet{brown70} for the He~{\sc i} series and by \citet{ferland80} for the He~{\sc ii} series. The continuum emissivity of heavy elements is
also calculated using equation~\ref{eq:milne}. In the hydrogenic case (i.e. H~{\sc i} and He~{\sc ii}), the two-photon continuum is calculated using the formalism described by \citet{nussbaumer84}; the data of \citet{drake69} are used for He~{\sc i}. Recombination lines
between lower levels n=2 through 8 and upper levels n=3 through 15, for H~{\sc i}, and lower levels
n=2 through 16 and upper levels n=3 through 30 for He~{\sc ii}, are calculated as a function of temperature according to the case~B data published by \citet{storey95}. The He~{\sc i} recombination lines are calculated as a function of temperature using the data of \citet{benjamin99}. In general, He~{\sc I} singlet lines follow Case~B whereas triplet lines follow Case~A (as there is no n~=~1 level for the triplets). Transitions to the 1$^1$S ground state of He~{\sc i} produce lines
which are capable of ionizing H and low ionization stages of higher elements. In particular,
the emissivities of the He~{\sc i} Lyman lines from n=2 through n=5 \citep{brocklehurst72} and the
intercombination lines corresponding to the transitions 2$^3$S-1$^1$S  and 2$^3$P-1$^1$S are
estimated as a function of temperature using the data of \citet{robbins68}. The contributions due to these lines to the total
energy distribution, from which the probability density functions are derived, are added into the respective energy bins.
Similarly, He~{\sc ii} Lyman lines can ionize both neutral hydrogen and neutral helium, as well as some of the low ions of heavier elements. Therefore the emissivities of He~{\sc ii} Lyman
lines with upper levels from n=2 through n=5 \citep[fits to][]{storey95} are also estimated as a function of temperature 
and their contributions to
the total energy distribution added into the respective frequency bin, as for the He~{\sc i}
lines. This method is based on the fact that all emission profiles are currently treated as
$\delta$ functions and  the line opacity is assumed to be zero;  and the absorption of energy packets is only due to the continuum opacity.  
 Finally, the emissivities of the collisional lines of the heavier ions are
calculated. This is done by using matrix inversion procedures in order to calculate the level populations of the ions. Appendix~1 contains references for the atomic data 
used for each ion. 

The energy distribution is derived from the total emissivity, summing over all the
contributions in a particular frequency interval.  The non-ionizing line emission  is treated separately, since, whenever such line packets are created,  they escape without
further interaction \footnote{Resonance lines longward of 912~{\AA} (e.g. C~{\sc iv}$\lambda\lambda$1548, 1550) may, in fact, diffuse out of the nebula via resonant scattering and may also be absorbed by dust during such diffusion. A treatment of dust grains will be included in future developments of the Mocassin code, and such effects may then be accounted for.}.

Once the line and continuum emissivities have been calculated, the probability that the
absorption of an ionizing energy packet will be followed by the emission of a non-ionizing packet is given by:
\begin{equation}
P_{esc} = \frac{\sum_{i}j^{l}_{{\rm X}^i}+\int_{0}^{\nu_H}j^{c}_{\nu} d\nu}
{\sum_{i}j^{l}_{{\rm X}^i}+\sum j^{l}_{\rm HeI}+\sum j^{l}_{\rm HeII}+ \int_{0}^{\nu_{max}}j^{c}_{\nu}d\nu}
\end{equation}
where $\nu_{\rm max}$ is the higher limit of the frequency grid; the $j^{l}_{{\rm X}^i}$ are the emissivities of the non-ionizing recombination lines of all
species considered; $j^{c}_{\nu}$ is the frequency dependent continuum emissivity;
$j^{l}_{{\rm HeI}}$
and $j^{l}_{{\rm HeII}}$ are the contributions due to those recombination lines of He~{\sc i} and He~{\sc ii}
which
are capable of ionizing neutral hydrogen and neutral helium. 
The choice between the re-emission of an ionizing photon or a non-ionizing one is made at this
point in the code. 

If an ionizing energy packet is to be re-emitted, then the new frequency
will be calculated according to the normalised cumulative probability density function for the ionizing
radiation, given by
\begin{equation}
p(\nu) = \frac{\int_{\nu_{\rm H}}^{\nu}j^{c}_{\nu'}d\nu'+\sum j^{l}_{\rm HeI}+\sum j^{l}_{\rm HeII}} 
{\int_{\nu_{\rm H}}^{\nu_{max}}j^{c}_{\nu'}d\nu'+\sum j^{l}_{\rm HeI}+\sum j^{l}_{\rm HeII}}
\label{eq:pdf}
\end{equation}
where, as usual,  the contributions due to the He~{\sc i} and He~{\sc ii} lines are added in the
corresponding frequency bins. If a non-ionizing energy packet is to be re-emitted, then its frequency must be determined from the probability density function for non-ionizing radiative energy, which is analogous to equation~\ref{eq:pdf}.

\subsection{The Iterative Procedure}
An initial guess of the physical conditions in the nebular cells, such as the ionization structure, electron temperature and electron density, needs to be specified before 
the simulation can begin. Procedures in Mocassin have been constructed such that only an initial guess at the electron temperature (which is initially set to a constant value throughout the nebula) must be included in the input file. Mocassin can then guess an initial ionization structure and, hence, the electron density. However, if the output of a one dimensional model (or a combination of more than one of them) is available, there are also procedures built into Mocassin to map these onto the three-dimensional Cartesian grid, by using simple interpolation routines. A one-dimensional mode option was implemented in Mocassin for this purpose. Several tests have shown that while the choice of the initial conditions has, of course, no influence on the final result of the simulation, it can, however, have an inpact on the number of iterations required to reach convergence. It is hard to quantify the number of iterations required for
convergence by each method, in particular, it depends strongly on the initial temperature input used in the first method, and, when applying the second method, on the deviation of the actual three-dimensional geometry from the simplified one-dimensional model used. However, with sufficient energy packets, the benchmark models described here should be fully converged in approximately fifteen to twenty iterations. A strategy to speed up the simulations is described in Section~3.1.

Once the initial conditions are specified, the frequency dependent total emissivities are calculated in each grid cell in order to set up the probability density functions for re-emitted radiation,  which are used for the determination of the frequency distribution of the re-emitted energy-packets during the Monte Carlo simulation.  The energy packets are then fired through the grid and their trajectories computed. Once all the energy packet trajectories have been computed, the Monte Carlo estimators for the mean intensity of the stellar and the diffuse radiation fields can be obtained, as described in Section~\ref{sub:meanIntensity}. The ionization fraction and the electron temperatures and densities must now be updated to be self-consistent with the current estimates of the radiation field at each grid point. This means solving the local ionization balance and thermal equilibrium equations simultaneously. The entire procedure is repeated until convergenge is achieved. The convergence criterion that is used in this work is based on the change of the local hydrogen ionization structure between successive iterations. In some cases, however, this is not a suitable convergence criterion (e.g. in hydrogen-deficient environments), for this reason, other criteria are also implemented in the code (e.g. based on the change of the local helium ionization structure, or of the local electron temperature between successive iterations), and these can be easily selected by using the appropriate switches in the input file. 

\subsection{Comparison of the Model with Observations}
\label{sub:comparison}
When the model has converged to its final solution, the output spectrum can be computed and compared with the results obtained from other models or with observational data. The total luminosity of the nebula emitted in various emission lines longward of the Lyman limit can be obtained by using two methods. The first method, which is only available to Monte Carlo codes, consists of summing up the number of energy packets in the given   
line, $N_{line}$, over the grid cells. Hence, the power emitted in the line is
given by
\begin{equation}
L_{line} = \frac{\varepsilon_0}{\Delta t}\sum_{i=1}^{i_{max}}\sum_{j=1}^{j_{max}}\sum_{k=1}^{k_{max}}N_{line}(x_i,y_j,z_k)
\label{eq:lineLuminosity}
\end{equation}
where $\frac{\varepsilon_0}{\Delta t}$ is given by equation~\ref{eq:enpacket}. The second method consists of using the values of the local electron temperature and ionic abundances given by the final converged  model solution to obtain the line emissivities for each grid cell. The luminosity of the nebula in any given line can then be calculated easily by summing the emissivity of the required line over the volume of the nebula. 

A comparison of the results obtained using the two methods described above, provides an indication of the level of accuracy achieved during the simulation, as the two methods will give consistent results only if enough energy packets have been used to yield good statistics for every line. In general, the second method (formal solution) yields the most accurate results, particularly for weak lines, which may emit relatively few photons. For the benchmark cases presented here, reasonable accuracy was deemed to have been achieved when the fluxes of the strongest transitions obtained using the pure Monte Carlo method were within 10\% of those obtained using the formal solution. Both methods can also be used to calculate line of sight results and to simulate long-slit observations. However, just as for the calculation of the integrated spectrum, the formal solution method is to be preferred, as it yields the most accurate results, particularly for the weaker lines.

In addition to the integrated emergent spectrum, other useful comparisons with the observations can be carried out, e.g. projected images of the final model nebula in a given line or at a given continuum frequency can be produced for arbitrary viewing angles. These can be compared directly with nebular images obtained in an appropriate filter. Mocassin computes and stores the physical properties of the nebula, as well as the emissivities of the gas at each grid point; these can be fed into IDL plotting routines in order to produce maps (Morisset et al., 2000). Also, by assuming a velocity field, line spectral profiles can be produced, together with position-velocity diagrams.  These can be compared with observations, if available, to deduce spatio-kinematic information about the object being studied. More information about the original IDL routines is given by \citet{morisset00} and \citet{monteiro00}. Details of the actual application to Mocassin's grid files are available in a companion paper on the modelling of the planetary nebula NGC~3918 \citep[][Paper~{\sc ii}]{ercolanoII}.

At the end of each Monte Carlo iteration the physical quantities which characterise the grid are written out to disc into three files, namely {\it grid1.out}, {\it grid2.out} and {\it grid3.out}. The first file contains the local electron temperature and density as well as the gas density at each grid cell, the second the ionization structure of the nebula and the third a number of model parameters, including the number of energy packet to be used in the simulation. These files are used in conjuction with a {\it warm start} driver, which allows an interrupted simulation to be resumed from the end of the last Monte Carlo simulation. This means that once a simulation has been interrupted the number of energy packets used (and indeed other model parameters, if required) can be adjusted, before the simulation is restarted, by modifying the file {\it grid3.out}. This feature can be used to speed up the simulations by using the following approach. The first few iterations are run using a lower number of energy packets than actually needed; so, for example, if the optimum number of energy packets for a given model is 10$^6$, then the first few iterations can be carried out using only 10$^5$ packets, hence reducing the run time for these by a factor of ten. This will result in about 50\%-60\% of the grid cells converging; in general, the inner cells converge more quickly, due to the larger number of sampling units available there (due mainly to geometrical dilution and to the reprocessing of energy packets to non-ionizing energy packets). At this point the simulation is interrupted and then resumed, after having adjusted the number of energy packets to the final required value (i.e. 10$^6$, in the previous example). Final convergence will then be achieved, in most cases, within four or five further iterations. The actual number of iterations required depends on the number of energy packets used: the larger the number of sampling quanta available at each cell, the quicker the cells will converge to a solution. The numbers quoted above, however, also depend on each particular model's geometry and optical thickness. 

\subsection{General Architecture}
\label{sec:mocassinArchitecture}

The Mocassin code was written using the Fortran~90 programming language. The code was developed and run initially on a Compaq(Dec) XP1000 with a 500~MHz CPU and  1~Gb
of memory and a preliminary serial version of the code still exists. A fully parallel version of the code has since been developed using Multiple Processes Interface (MPI) routines and it currently runs on a Silicon Graphics Origin 2000 machine with 24 processors and 6~Gb of memory and a SUN Microsystems Sunfire V880 machine with 16 processors and 64 Gb of memory. Monte Carlo simulations are, by their nature, very parallelizable problems and, indeed, Mocassin can achieve a linear {\it speed-up}, i.e. a speed-up that is directly proportional to the number of processors used.  A detailed description of all the Mocassin modules, input commands and output files is given by \citet[][PhD Thesis]{ercolano02}. A copy of the code is available from the author (be@star.ucl.ac.uk) together with the relevant thesis chapters. 

\section{Application to benchmark cases}
\label{sec:benchmarks}

Numerical simulations of photoionized nebulae are very complex and a number of factors, such as numerical approximations and assumptions, and the complexity of the calculation 
itself, introduce a degree of uncertainty into the results. For this reason, it is important for modelers to have certain standards of comparison, in order to identify problems in their codes and to reach an adequate degree of accuracy in their calculation. A series of meeting have been held, beginning in Meudon, France, in 1985 \citep{pequignot86} and in Lexington, Kentucky, first in 1995 \citep{ferland95} and again in 2000 \citep{pequignot01}, in order to define a set of benchmark 
cases which could be used by all photoionization modelers to test their codes against. The benchmarks which resulted from these meetings include H~{\sc ii}~regions, planetary 
nebulae, narrow line regions (NLRs) of AGNs and X-ray slabs. Mocassin does not have, at present, the capability to treat NLRs and X-ray slabs, as some relevant physical 
processes, such as Compton heating and inner shell ionization, are not yet included. For this reason, only the H~{\sc ii}~region and planetary nebula benchmarks are performed in this work. The expansion of the code to include high energy processes is planned in the future.

Results from several other codes are available for comparison; these are all one-dimensional codes and, apart from differences in the atomic data used by each of them, their main differences lie in the treatment of the diffuse radiation field transfer. A brief description of each of these codes is given by \citet{ferland95}. 
Although the majority of these codes have evolved somewhat since the 1995 Lexington meeting, mostly via the updating of the atomic data sets and the inclusion of more and 
specialised physical processes, their basic structures have stayed the same. The seven codes included for comparison are 
G. Ferland's {\it Cloudy} (GF), J.P Harrington's code (PH), D. P\'equignot's {\it Nebu} (DP), T. Kallman's {\it XStar} (TK), H. Netzer's {\it Ion} (HN), R. Sutherland's {\it Mappings} (RS) 
and R. Rubin's {\it Nebula} (RR). Only two of these codes,  the Harrington code and Rubin's {\it Nebula}, treat the diffuse radiative transfer exactly. 
The others use some versions of the {\it outward-only approximation} of varying sophistication. In this approximation all diffuse radiation is assumed to be emitted isotropically into the outward half of space. 

\begin{table}
\caption{Lexington 2000 benchmark model input parameters. }
\label{tab:inputpar}
\begin{center}
\begin{tabular}{lcccc}
\hline
Parameter 	& HII40	& HII20	& PN150	& PN75	 \\
\hline
L(BB)/10$^{37}$($\frac{erg}{sec}$) & 308.2 & 600.5 & 3.607 & 1.913 \\
T(BB)/10$^3$K	& 40	& 20	& 150 	& 75	\\
$R_{\rm in}$/10$^{17}$cm & 30	& 30	& 1 	& 1.5	\\
$n_{\rm H}$/cm$^{-3}$	& 100	& 100	& 3000	& 500	\\
He/H		& 0.10	& 0.10	& 0.10 	& 0.10	\\
C/H$\times$ 10$^5$	& 22.	& 22.	& 30.	& 20.	\\
N/H$\times$ 10$^5$	& 4. 	& 4.	& 10.	& 6.	 \\
O/H$\times$ 10$^5$	& 33.	& 33.	& 60. 	& 30.	\\
Ne/H$\times$ 10$^5$	& 5.	& 5.	& 15. 	& 6. 	\\
Mg/H$\times$ 10$^5$	& -	& -	& 3. 	& 1.	\\
Si/H$\times$ 10$^5$	& -	& - 	& 3. 	& 1.	\\
S/H$\times$ 10$^5$ 	& 0.9	& 0.9	& 1.5 	& 1.	\\
\hline
\multicolumn{5}{c}{} \\
\end{tabular}
\\
\small{Elemental abundances are by number with respect to H.}
\end{center}
\end{table}

The predicted line fluxes from each code for each benchmark case are listed in Tables~\ref{tab:hii40} to~\ref{tab:pn75}, 
together with the volume-averaged mean electron temperature, weighted by the proton and electron densities, $N_{\rm p}N_{\rm e}$, $<$T$[$N$_{\rm p},$N$_{\rm e}]>$, the electron temperature at the inner edge of the nebula, 
T$_{inner}$, and the mean ratio of fractional He$^+$ to fractional H$^+$, $\frac{<He^+>}{<H^+>}$, which represents the fraction of helium in the H$^+$ region that is singly ionized. $<$T$[$N$_{H^+},$N$_{\rm e}]>$ and $\frac{<He^+>}{<H^+>}$ are calculated according to the following equations \citep{ferland95}

\begin{equation}
<T[N_{\rm p},N_{\rm e}]> = \frac{\int N_{\rm e} N_{\rm p} T_{\rm e} \, dV}{\int N_{\rm e} N_{\rm p} \, dV}
\label{eq:meantempferl}
\end{equation}

and

\begin{equation}
\frac{<{\rm He}^+>}{<{\rm H}^+>} = \frac{n({\rm H})}{n({\rm He})} \frac{\int N_{\rm e}N_{{\rm He}^+} \, dV}{\int N_{\rm e} N_{\rm p} \, dV}
\label{eq:meanionferl}
\end{equation}

where $N_{\rm e}$ and $N_{\rm p}$ are the local electron and proton densities, respectively, $N_{{\rm He}^+}$ is the density of He$^+$, and $n(H)$ and $n$(He) are the total hydrogen and helium densities. 

Table~\ref{tab:inputpar} lists the input parameters for all the benchmark models dicussed here. All the benchmark cases listed in Table~\ref{tab:inputpar} were calculated using both the three-dimensional and the one-dimensional mode of Mocassin and both sets of results are included here for comparison. It is clear from Tables~\ref{tab:hii40} to~\ref{tab:pn75} that the results of the three-dimensional and one-dimensional modes of Mocassin are consistent with each other. The small differences that do exist can be entirely attributed to the coarseness of the grids used for the three-dimensional calculations. The aim of the benchmarking described in this work is to assess the reliability of Mocassin in its fully {\it three-dimensional} mode, for this reason the one-dimensional mode results will not be included in the following performance analysis; moreover the inclusion of two sets of results from what is, essentially, the same code would introduce a bias in the median and isolation factors calculations described below. Finally, to avoid any confusion, any mention of Mocassin throughout the rest of this paper refers to the fully three-dimensional version of the code, unless otherwisestated. 

Figures~\ref{fig:xhii} and \ref{fig:xpn} show the electron temperatures (top panels) and the fractional ionic abundances of oxygen (middle panels) and carbon (bottom panels) for the four benchmark cases analysed. The ionic abundances in every cell in the ionized region are plotted against radial distance from the star. These plots are interesting not only because they provide a clear picture of the overall temperature and ionization structure of each model, but also because from the scatter of the data points one can estimate the accuracy of the final results. (Note that such plots are only meaningful in the spherically symmetric case.) 

Four benchmark model nebulae were computed, two H~{\sc ii} regions and two planetary nebulae. These benchmarks were designed to be uncomplicated yet to test different aspects of the modelling \citep[see][]{ferland95}. The nebulae are homogeneous in density and, for simplicity, blackbodies are used as the ionizing sources instead of model stellar atmopheres.

Following the analysis of P\'equignot \citep[see][]{pequignot01}, isolation factors, $if$'s, were computed for each predicted quantity in each case study. These are defined as the ratio of the largest to the 
penultimate largest value of a given output quantity or the ratio of the penultimate smallest value to the smallest value. These ratios are computed with the intention to identify aberrant values. 
A large $if$ can be attributed to a number of factors, but often these can be attributed to a difference in the atomic data used by each modeler. 
A list of the number of $if$'s larger than 1.01, 1.03, 1.10, 1.30 and 2.00 is given in Table~\ref{tab:ifs}, for each benchmark. After analysing the benchmark results obtained by all the modelers who participated in the Lexington workshop, \citet{pequignot01} suggested that an isolation factor larger than 1.30 is indicative of a significant departure and a possible problem. A large number of occurences of $if'{\rm s}~>~1.30$ should either have an acceptable explanation or lead to corrections to the code. 

The number of results not predicted by any given code is listed in the {\it No pred} row of Table~\ref{tab:ifs}. \citet{pequignot01} also noted, in the proceedings of the November 2000 Lexington meeting, that the lack of a prediction for a particular observable may simply reflect a lack of interest by the modeller in it; on the other hand, a frequent occurence of {\it No pred} may also indicate limitations in the predictive power of a given code. 

As argued by \citet{pequignot01}, a large error can be introduced when the average over a small sample containing a number of aberrant values is taken. In order to minimise this error, median values are calculated instead of averages and these are given for each observable listed in Tables~\ref{tab:hii40} to \ref{tab:pn75}, in the column labelled {\it Med}. The medians are calculated to the precision shown in Tables~\ref{tab:hii40} to \ref{tab:pn75}. Table~\ref{tab:medians} lists the number of median values scored by each code for each benchmark, i.e. the number of times the code was the closest to the median value. When a median value is shared by two or more codes the score is given to each one, therefore the sum of the median values scored by all the codes is higher than the number of observables (the column labelled {\it Total} in Table~\ref{tab:medians}).  

\begin{table}
\begin{center}
\caption{Summary of the number of energy packets needed for $>$\,50\% and $>$\,95\% convergence (see text for explanation) for each of the benchmark cases}
\begin{tabular}{lcccccc}
\multicolumn{7}{c}{} \\
\hline
Case			& \multicolumn{3}{c}{$\tau_{edge}$} & $n_x{\times}n_y{\times}n_z$ & \multicolumn{2}{c}{$N_{\rm packets}$} \\
				& H$^0$	& He$^0$& He$^+$& 			&  $>$50\% 	  & $>$95\% \\
\hline
HII40				& 4.79	& 1.15	& 177.8	& 13$\times$13$\times$13 & 5$\cdot$10$^5$ & 5$\cdot$10$^6$ \\
HII20				& 2.95	& 1.13	& 91.2	& 13$\times$13$\times$13 & 5$\cdot$10$^6$ & 5$\cdot$10$^7$ \\
PN150				& 34.0	& 6.87	& 57.9	& 13$\times$13$\times$13 & 3$\cdot$10$^5$ & 3$\cdot$10$^6$ \\
PN75				& 1.16	& 0.24	& 31.5	& 13$\times$13$\times$13 & 4$\cdot$10$^5$ & 4$\cdot$10$^6$ \\
\hline
\end{tabular}
\label{tab:nphots}
\end{center}
\end{table}

\subsection{Sampling Requirements}
Table~\ref{tab:nphots} lists the optical depths at the ionization threshold frequencies for H$^0$, He$^0$ and He$^+$, at the outer edge of the grids, for the four benchmark models analyzed here. For each model, the number of grid points is also given (column 5), together with the number of energy packets used, $N_{\rm packets}$, according to the two-step stategy described above, first to achieve convergence in 50\%-60\% of the total number of grid cells ($>50\%$, column 6) and then to achieve total convergence ($>95\%$, column 7). Table~\ref{tab:nphots} shows that the softer the ionizing radiation field, the larger the number of energy packets required to achieve a given degree of convergence. The reason for this effect is that in a softer radiation field case the number of energy packets emitted at wavelengths shorter than the Lyman limit will be less than in the case of a harder radiation field. A larger total number of energy packets then needs to be used in order to obtain a number of ionizing photons adequate to properly sample the nebula. The aim of Table~\ref{tab:nphots} is merely to provide some general guidelines for selecting the appropriate number of energy packets for a particular simulation; however, as stated before, the optimum number should be determined for each given model, particularly in non-spherically symmetric cases. 
 
\subsection{Benchmark Results}
\begin{table}
\begin{center}
\caption{Deviation of the Monte Carlo method from the formal solution for the prediction of some significant line fluxes in the benchmark models.}
\begin{tabular}{lcccc}
\multicolumn{5}{c}{} \\
\hline
Line 			& HII40 & HII20  & PN150  & PN75 \\
\hline
H$\beta$		& 2.7\% & 9.5\%  & 5.8\%  & 2.8\% \\ 
He~{\sc i}~5876~{\AA}	& 5.2\%	& 6.3\%	 & 0.96\% & 4.5\% \\
$[$N~{\sc ii}$]$~6584~{\AA} & 7.6\%	& 4.9\%  & 8.5\%  & 4.8\% \\
$[$O~{\sc ii}$]$~5007~{\AA}	& 3.1\% & 12.0\% & 4.0\%  & 1.1\% \\	 
$[$S~{\sc iii}$]$~9532~{\AA}& 5.8\%	& 5.0\%	 & 2.0\%  & 2.0\% \\
\hline
\end{tabular}
\label{tab:anMon}
\end{center}
\end{table}

\renewcommand{\baselinestretch}{1.2}
\begin{table*}
\begin{center}
\begin{minipage}{13.5cm}
\caption{Lexington 2000 Standard H~{\sc ii} region (HII40) benchmark case reults. }
\label{tab:hii40}
\begin{tabular}{lr|ccccccccc}
\hline
Line				& Median & GF	& HN	& DP	& TK	& PH	& RS	& RR	&\multicolumn{2}{c}{BE}\\
				&	&	&	&	&	&	&	& 	& 3-D	& 1-D \\
\hline
H$\beta$/10$^{37}$ erg/s	& 2.05 	& 2.06	& 2.02	& 2.02	& 2.10	& 2.05	& 2.07	& 2.05 	& 2.02	& 2.10 \\
H$\beta$ 4861			& -	& 1.00	& 1.00	& 1.00	& 1.00 	& 1.00 	& 1.00	& 1.00 	& 1.00	& 1.00 \\
He~{\sc i} 5876     		& 0.116	& 0.119	& 0.112	& 0.113	& 0.116 & 0.118	& 0.116	& -	& 0.114	& 0.112 \\
 C~{\sc ii}$]$ 2325+		& 0.144	& 0.157	& 0.141	& 0.139	& 0.110	& 0.166 & 0.096	& 0.178	& 0.148	& 0.126 \\
 C~{\sc ii} 1335		& 0.082	& 0.100	& 0.078	& 0.094	& 0.004	& 0.085	& 0.010	& - 	& 0.082 & 0.084 \\
 C~{\sc iii}$]$ 1907+1909 	& 0.070	& 0.071	& 0.076	& 0.069	& 0.091	& 0.060	& 0.066	& 0.074	& 0.041 & 0.041 \\
$[$N~{\sc ii}$]$ 122.$\mu$m	& 0.034	& 0.027	& 0.037	& 0.034	& - 	& 0.032	& 0.035	& 0.030	& 0.036	& 0.034 \\
$[$N~{\sc ii}$]$ 6584+6548	& 0.730	& 0.669	& 0.817 & 0.725	& 0.69 	& 0.736 & 0.723 & 0.807	& 0.852	& 0.786 \\
$[$N~{\sc ii}$]$ 5755		& .0054	& .0050	& .0054	& .0050	& - 	& .0064	& .0050	& .0068	& .0061 & .0054 \\
$[$N~{\sc iii}$]$ 57.3 $\mu$m	& 0.292	& 0.306	& 0.261	& 0.311	& - 	& 0.292	& 0.273	& 0.301	& 0.223	& 0.229 \\
$[$O~{\sc i}$]$ 6300+6363	& .0086	& .0094	& .0086	& .0088	& .012	& .0059	& .0070	& - 	& .0065	& .0080 \\
$[$O~{\sc ii}$]$ 7320+7330	& 0.029	& 0.029	& 0.030	& 0.031	& 0.023	& 0.032	& 0.024	& 0.036	& 0.025	& 0.022 \\
$[$O~{\sc ii}$]$ 3726+3729	& 2.03	& 1.94	& 2.17	& 2.12	& 1.6	& 2.19	& 1.88	& 2.26	& 1.92	& 1.75 \\
$[$O~{\sc iii}$]$ 51.8 $\mu$m   & 1.06	& 1.23	& 1.04	& 1.03	& 0.99	& 1.09	& 1.06	& 1.08	& 1.06	& 1.09 \\
$[$O~{\sc iii}$]$ 88.3 $\mu$m   & 1.22	& 1.12	& 1.06	& 1.23	& 1.18	& 1.25	& 1.23	& 1.25	& 1.22	& 1.26 \\
$[$O~{\sc iii}$]$ 5007+4959	& 2.18	& 2.21	& 2.38	& 2.20	& 3.27	& 1.93	& 2.17	& 2.08	& 1.64	& 1.70 \\
$[$O~{\sc iii}$]$ 4363		& .0037	& .0035	& .0046	& .0041	& .0070	& .0032	& .0040	& .0035	& .0022	& .0023 \\
$[$O~{\sc iv}$]$ 25.9 $\mu$m	& .0010	& .0010	& .0010	& .0010	& .0013	& .0013	& .0010	& -	& .0010	& .0010 \\ 
$[$Ne~{\sc ii}$]$ 12.8 $\mu$m   & 0.195	& 0.177	& 0.195	& 0.192	& - 	& 0.181	& 0.217	& 0.196	& 0.212	& 0.209 \\
$[$Ne~{\sc iii}$]$ 15.5 $\mu$m  & 0.322	& 0.294	& 0.264	& 0.270	& 0.35	& 0.429	& 0.350	& 0.417	& 0.267	& 0.269 \\
$[$Ne~{\sc iii}$]$ 3869+3968	& 0.085	& 0.084	& 0.087	& 0.071	& 0.092	& 0.087	& 0.083	& 0.086	& 0.053	& 0.055 \\
$[$S~{\sc ii}$]$ 6716+6731	& 0.147	& 0.137	& 0.166	& 0.153	& 0.315	& 0.155	& 0.133	& 0.130	& 0.141	& 0.138 \\
$[$S~{\sc ii}$]$ 4068+4076	& .0080	& .0093	& .0090	& .0100	& .026	& .0070	& .005	& .0060	& .0060	& .0057 \\
$[$S~{\sc iii}$]$ 18.7 $\mu$m	& 0.577	& 0.627	& 0.750	& 0.726	& 0.535	& 0.556	& 0.567	& 0.580	& 0.574	& 0.569 \\
$[$S~{\sc iii}$]$ 33.6 $\mu$m	& 0.937	& 1.24	& 1.43	& 1.36	& 0.86	& 0.892	& 0.910	& 0.936	& 0.938	& 0.932 \\
$[$S~{\sc iii}$]$ 9532+9069	& 1.22	& 1.13	& 1.19	& 1.16	& 1.25	& 1.23	& 1.25	& 1.28	& 1.21	& 1.19 \\
$[$S~{\sc iv}$]$ 10.5 $\mu$m	& 0.359	& 0.176	& 0.152	& 0.185	& 0.56	& 0.416	& 0.388	& 0.330	& 0.533	& 0.539 \\
10$^3$\,$\times$\,$\Delta$(BC 3645)/$\AA$& 5.00	& 4.88	& - & 4.95 & -	& 5.00	& 5.70	& - 	& 5.47	& 5.45 \\
$T_{\rm inner}$/ K   		& 7653	& 7719	& 7668	& 7663	& 8318	& 7440	& 7644	& 7399	& 7370	& 7480 \\	
$<$T$[$N$_{\rm p}$N$_{\rm e}]>$/K 	& 8026	& 7940	& 7936	& 8082	& 8199	& 8030	& 8022	& 8060	& 7720	& 7722 \\
$R_{\rm out}$/10$^{19}$cm		& 1.46	& 1.46	& 1.46	& 1.46	& 1.45	& 1.46	& 1.47	& 1.46	& 1.46	& 1.49 \\
$<$He$^+>/<$H$^+>$		  	& 0.767 & 0.787 & 0.727 & 0.754 & 0.77	& 0.764	& 0.804	& 0.829	& 0.715 & 0.686 \\
\hline
\end{tabular}
\small{GF: G. Ferland's {\it Cloudy}; PH: J.P Harrington code; DP: D. P\'equignot's {\it Nebu}; TK: T. Kallman's {\it XStar}; HN: H. Netzer's {\it Ion}; RS: R. Sutherland's {\it Mappings}; RR: R. Rubin's {\it Nebula}; BE: B. Ercolano's {Mocassin}.}
\end{minipage}
\end{center}
\end{table*}
\renewcommand{\baselinestretch}{1.5}

\begin{table*}
\begin{center}
\begin{minipage}{13.5cm}
\caption{Lexington 2000 low excitation H~{\sc ii} region (HII20) Benchmark case results. }
\begin{tabular}{lr|ccccccccc}
\hline
Line				& Med & GF	& HN	& DP	& TK	& PH	& RS	& RR	& \multicolumn{2}{c}{BE} \\
				&	&	&	&	&	&	&	&	& 3-D	& 1-D \\
\hline
H$\beta$/10$^{36}$ erg/s	& 4.91	& 4.85	& 4.85	& 4.83	& 4.9	& 4.93	& 5.04	& 4.89	& 4.97 & 5.09 \\
H$\beta$ 4861			& -	& 1.00	& 1.00	& 1.00	& 1.00 	& 1.00 	& 1.00	& 1.00 	& 1.00 & 1.00 \\
He~{\sc i} 5876     		& .0074	& .0072	& 0.008	& .0073	& 0.008	& .0074	& .0110	& -	& .0065 & .0074 \\
 C~{\sc ii}$]$ 2325+		& 0.046	& 0.054	& 0.047	& 0.046	& 0.040	& 0.060	& 0.038	& 0.063	& 0.042 & 0.031 \\
$[$N~{\sc ii}$]$ 122.$\mu$m	& 0.071	& 0.068	& -	& 0.072	& 0.007	& 0.072	& 0.071	& 0.071	& 0.071 & 0.070 \\
$[$N~{\sc ii}$]$ 6584+6548	& 0.823	& 0.745	& 0.786	& 0.785	& 0.925	& 0.843	& 0.803	& 0.915	& 0.846 & 0.771 \\
$[$N~{\sc ii}$]$ 5755		& .0028	& .0028	& .0024	& .0023	& .0029	& .0033	& .0030	& .0033	& .0025 & .0021 \\
$[$N~{\sc iii}$]$ 57.3 $\mu$m	& .0030	& .0040	& .0030	& .0032	& .0047	& .0031	& .0020	& .0022	& .0019 & .0032 \\
$[$O~{\sc i}$]$ 6300+6363	& .0060	& .0080	& .0060	& .0063	& .0059	& .0047	& .0050	& -	& .0088 & .0015 \\
$[$O~{\sc ii}$]$ 7320+7330	& .0086	& .0087	& .0085	& .0089	& .0037	& .0103	& .0080	& .0100	& .0064 & .0051 \\
$[$O~{\sc ii}$]$ 3726+3729	& 1.09	& 1.01	& 1.13	& 1.10	& 1.04	& 1.22	& 1.08	& 1.17	& 0.909 & 0.801 \\
$[$O~{\sc iii}$]$ 51.8 $\mu$m   & .0012	& .0014	& .0012	& .0012	& .0016	& .0013	& .0010	& .0008	& .0010 & .0011 \\
$[$O~{\sc iii}$]$ 88.3 $\mu$m 	& .0014 & .0016	& .0014	& .0014	& .0024	& .0014	& .0010	& .0009	& .0012 & .0013 \\
$[$O~{\sc iii}$]$ 5007+4959	& .0014	& .0021	& .0016	& .0015	& .0024	& .0014	& .0010	& .0010	& .0011 & .0012 \\
$[$Ne~{\sc ii}$]$ 12.8 $\mu$m 	& 0.273	& 0.264	& 0.267	& 0.276	& 0.27	& 0.271	& 0.286	& 0.290	& 0.295 & 0.296 \\
$[$S~{\sc ii}$]$ 6716+6731	& 0.489	& 0.499	& 0.473	& 0.459	& 1.02	& 0.555	& 0.435	& 0.492	& 0.486 & 0.345 \\
$[$S~{\sc ii}$]$ 4068+4076	& 0.017	& 0.022	& 0.017	& 0.020	& 0.052	& 0.017	& 0.012	& 0.015	& 0.013 & .0082 \\
$[$S~{\sc iii}$]$ 18.7 $\mu$m	& 0.386	& 0.445	& 0.460	& 0.441	& 0.34	& 0.365	& 0.398	& 0.374	& 0.371 & 0.413 \\
$[$S~{\sc iii}$]$ 33.6 $\mu$m	& 0.658	& 0.912	& 0.928	& 0.845	& 0.58	& 0.601	& 0.655	& 0.622	& 0.630 & 0.702 \\
$[$S~{\sc iii}$]$ 9532+9069	& 0.537	& 0.501	& 0.480	& 0.465	& 0.56	& 0.549	& 0.604	& 0.551	& 0.526 & .582 \\
10$^3$\,$\times$\,$\Delta$(BC 3645)/$\AA$& 5.57	& 5.54	& -	& 5.62	& -	& 5.57	& 5.50	& - & 6.18 & 6.15 \\
$T_{\rm inner}$/ K   		& 6765	& 7224	& 6815	& 6789	& 6607	& 6742	& 6900	& 6708	& 6562 & 6662 \\
$<$T$[$N$_{\rm p}$N$_{\rm e}]>$/K 	& 6662	& 6680	& 6650	& 6626	& 6662	& 6749	& 6663	& 6679	& 6402 & 6287 \\
$R_{\rm out}$/10$^{18}$cm		& 8.89	& 8.89	& 8.88	& 8.88	& 8.7	& 8.95	& 9.01	& 8.92	& 8.89 & 8.92 \\
$<$He$^+>/<$H$^+>$			& 0.048	& 0.048 & 0.051 & 0.049 & 0.048	& 0.044	& 0.077	& 0.034	& 0.041 & 0.048 \\ 
\hline
\end{tabular}
\small{GF: G. Ferland's {\it Cloudy}; PH: J.P Harrington code; DP: D. P\'equignot's {\it Nebu}; TK: T. Kallman's {\it XStar}; HN: H. Netzer's {\it Ion}; RS: R. Sutherland's {\it Mappings}; RR: R. Rubin's {\it Nebula}; BE: B. Ercolano's {Mocassin}.}
\end{minipage}
\end{center}
\label{tab:hii20}
\end{table*}

\renewcommand{\baselinestretch}{0.9}
\begin{table*}
\begin{center}
\begin{minipage}{12.5cm}
\caption{Lexington 2000 thick planetary nebula  (PN150) benchmark case results. }
\begin{tabular}{lr|cccccccc}
\hline
Line				& Med	& GF	& HN	& DP	& TK	& PH	& RS	&\multicolumn{2}{c}{BE}\\
				&	&	&	&	&	&	&	& 3-D	& 1-D \\
\hline
H$_{\beta}$/10$^{35}$ erg/s	& 2.79	& 2.86	& 2.83	&2.84	& 2.47	& 2.68	& 2.64	& 2.79  & 2.89 \\
H$_{\beta}$ 4861		& -	& 1.00	& 1.00	& 1.00	& 1.00 	& 1.00 	& 1.00	& 1.00  & 1.00 \\
He~{\sc i} 5876     		& 0.104 & 0.110	& 0.129	& 0.118	& 0.096	& 0.096	& 0.095	& 0.104 & 1.06 \\ 
He~{\sc ii} 4686		& 0.328 & 0.324	& 0.304	& 0.305	& 0.341	& 0.333	& - 	& 0.333 & 0.320 \\
C~{\sc ii}$]$ 2325+		& 0.293 & 0.277	& 0.277	& 0.293	& 0.346	& 0.450	& 0.141	& 0.339 & 0.330 \\
C~{\sc ii} 1335			& 0.119 & 0.121	& 0.116	& 0.130	& - 	& 0.119	& - 	& 0.103 & 0.104 \\
C~{\sc iii}$]$ 1907+1909 	& 0.174 & 1.68	& 1.74	& 1.86	& 1.69	& 1.74	& 1.89	& 1.72  & 1.71 \\
C~{\sc iv} 1549+		& 2.16	& 2.14	& 2.43	& 2.16	& 0.154	& 2.09	& 3.12	& 2.71  & 2.65 \\
$[$N~{\sc i}$]$ 5200+5198	& 0.012	& 0.013 & 0.022 & 0.010	& - 	& 0.020	& 0.005	& .0067 & 0.012 \\ 
$[$N~{\sc ii}$]$ 6584+6548	& 1.17	& 1.15	& 1.16	& 1.18	& 1.01	& 1.35	& 1.17	& 1.43  & 1.37 \\
$[$N~{\sc ii}$]$ 5755		& 0.017 & 0.017	& 0.016	& 0.017	& 0.020	& 0.023	& 0.016	& 0.022 & .0021 \\
 N~{\sc iii}$]$ 1749+		& 0.111 & 0.106	& 0.109	& 0.132	& 0.184	& 0.139	& 0.091	& 0.111 & 0.110 \\
$[$N~{\sc iii}$]$ 57.3 $\mu$m	& 0.129 & 0.129	& 0.133	& 0.134	& 0.12	& 0.135	& 0.126	& 0.120 & 0.122 \\ 
N~{\sc iv}$]$ 1487+		& 0.168 & 0.199	& 0.178	& 0.192	& 0.154	& 0.141	& 0.168	& 0.162 & 0.159 \\
 N~{\sc v} 1240+		& 0.147 & 0.147	& 0.159	& 0.154	& 0.055	& 0.107	& 0.248	& 0.147 & 0.145 \\
$[$O~{\sc i}$]$  63.1 $\mu$m	& 0.020 & 0.024	& 0.017	& 0.025	& -	& .0072	& 0.049	& 0.010 & 0.011 \\
$[$O~{\sc i}$]$ 6300+6363	& 0.135 & 0.144	& 0.126	& 0.135	& 0.245	& 0.104	& 0.101	& 0.163 & 0.153 \\
$[$O~{\sc ii}$]$ 3726+3729	& 2.11  & 2.03	& 1.96	& 2.32	& 2.11	& 2.66	& 1.75	& 2.24  & 2.25 \\
$[$O~{\sc iii}$]$ 51.8 $\mu$m 	& 1.39  & 1.30	& 1.45	& 1.42	& 0.954	& 1.39	& 1.28	& 1.50  & 1.52 \\
$[$O~{\sc iii}$]$ 88.3 $\mu$m   & 0.274 & 0.261	& 0.292	& 0.291	& 0.27	& 0.274	& 0.252	& 0.296 & 0.299 \\
$[$O~{\sc iii}$]$ 5007+4959	& 21.4  & 21.4	& 22.2	& 21.1	& 26.0	& 20.8	& 16.8	& 22.63 & 22.52 \\
$[$O~{\sc iii}$]$ 4363		& 0.155 & 0.152	& 0.151	& 0.156	& 0.249	& 0.155	& 0.109	& 0.169 & 0.166 \\
$[$O~{\sc iv}$]$ 25.9 $\mu$m	& 3.78  & 3.45	& 3.16	& 3.78	& 3.95	& 4.20	& 4.05	& 3.68  & 3.60 \\
O~{\sc iv}$]$ 1403+		& 2.30  & 0.183	& 0.236	& 0.324	& 0.357	& 0.225	& - 	& 0.203 & 0.201 \\
O~{\sc v}$]$ 1218+		& 0.169 & 0.165	& 0.189	& 0.170	& 0.142	& 0.097	& 0.213	& 0.169 & 0.168 \\
O~{\sc vi} 1034+		& 0.025 & 0.028	& 0.026	& 0.022	& 0.026	& 0.014	& - 	& 0.025 & 0.026 \\
$[$Ne~{\sc ii}$]$ 12.8 $\mu$m   & 0.030 & 0.028	& 0.032	& 0.030	& 0.020	& 0.027	& 0.043	& 0.030 & 0.031 \\
$[$Ne~{\sc iii}$]$ 15.5 $\mu$m  & 1.97  & 1.88	& 1.97	& 1.92	& 1.73	& 2.76	& 2.71	& 2.02  & 2.03 \\
$[$Ne~{\sc iii}$]$ 3869+3968	& 2.63  & 2.64	& 2.32	& 2.25	& 2.86	& 3.04	& 2.56	& 2.63  & 2.61 \\
$[$Ne~{\sc iv}$]$ 2423+		& 0.723	& 0.707 & 0.712	& 0.785	& 1.13	& 0.723	& 0.832	& 0.749 & 0.741 \\
$[$Ne~{\sc v}$]$ 3426+3346	& 0.692 & 0.721	& 0.706	& 0.661 & 1.07	& 0.583	& 0.591 & 0.692 & 0.687 \\
$[$Ne~{\sc v}$]$ 24.2 $\mu$m 	& 0.980	& 0.997	& 0.98	& 0.928	& 1.96	& 0.936	& 0.195	& 1.007 & 0.997 \\
$[$Ne~{\sc vi}$]$ 7.63 $\mu$m 	& 0.076	& 0.107	& 0.075	& 0.077	& 0.692	& 0.011 & - 	& 0.050 & 0.051 \\
Mg~{\sc ii} 2798+		& 1.22	& 2.22	& 2.10	& 1.22	& 0.023	& 0.555	& 0.863	& 2.32  & 2.32 \\
$[$Mg~{\sc iv}$]$ 4.49 $\mu$m 	& 0.111 & 0.121	& 0.111	& 0.107	& 0.13	& 0.042	& 0.115	& 0.111 & 0.109 \\
$[$Mg~{\sc v}$]$ 5.61 $\mu$m 	& 0.144	& 0.070	& 0.132	& 0.162	& 0.18	& 0.066	& - 	& 0.156 & 0.156 \\
$[$Si~{\sc ii}$]$ 34.8 $\mu$m	& 0.168	& 0.155	& 0.168	& 0.159	& 0.263	& 0.253	& 0.130	& 0.250 & 0.263\\
Si~{\sc ii}$]$ 2335+		& 0.159	& 0.160	& 0.155	& 0.158	& 0.20	& -	& 0.127 & 0.160 & 0.164 \\
Si~{\sc iii}$]$ 1892+		& 0.382	& 0.446	& 0.547	& 0.475	& 0.321	& 0.382	& 0.083	& 0.325 & 0.316 \\ 
Si~{\sc iv} 1397+		& 0.172	& 0.183	& 0.218	& 0.169	& 0.015	& 0.172	& 0.122	& 0.214 & 0.207 \\
$[$S~{\sc ii}$]$ 6716+6731	& 0.370	& 0.359	& 0.37	& 0.399	& 0.415	& 0.451	& 0.322	& 0.357 & 0.370 \\
$[$S~{\sc ii}$]$ 4069+4076	& 0.077	& 0.073	& 0.078	& 0.086	& 0.19	& 0.077	& 0.050	& 0.064 & 0.063 \\
$[$S~{\sc iii}$]$ 18.7 $\mu$m	& 0.578	& 0.713	& 0.788	& 0.728	& 0.15	& 0.488	& 0.578	& 0.495 & 0.505 \\
$[$S~{\sc iii}$]$ 33.6 $\mu$m	& 0.240	& 0.281	& 0.289	& 0.268	& 0.06	& 0.206	& 0.240	& 0.210 & 0.214  \\
$[$S~{\sc iii}$]$ 9532+9069	& 1.96	& 2.07	& 2.07	& 1.96	& 0.61	& 1.90	& 2.04	& 1.89  & 1.92 \\
$[$S~{\sc iv}$]$ 10.5 $\mu$m	& 2.22	& 2.09	& 1.65	& 1.76	& 2.59	& 2.22	& 2.25	& 2.25  & 2.22 \\
$T_{\rm inner}$/ K   		& 18100	& 18120	& 17950	& 18100	& 19050	& 17360	& 19100	& 16670 & 17703 \\
$<$T$[$N$_{\rm p}$N$_{\rm e}]>$/K 	& 12110	& 12080	& 13410	& 12060	& 13420	& 12110	& 11890	& 12150 & 12108 \\
$R_{\rm out}$/10$^{17}$cm		& 4.04	& 4.04	& 3.90	& 4.11	& 4.07	& 4.04	& 3.98	& 4.11  & 4.19 \\
$<$He$^+>/<$H$^+>$			& 0.704	& 0.702 & 0.726 & 0.714 & 0.79	& 0.696 & 0.652	& 0.702 & 0.711 \\
\hline
\end{tabular}
\small{GF: G. Ferland's {\it Cloudy}; PH: J.P Harrington code; DP: D. P\'equignot's {\it Nebu}; TK: T. Kallman's {\it XStar}; HN: H. Netzer's {\it Ion}; RS: R. Sutherland's {\it Mappings}; BE: B. Ercolano's {Mocassin}.}
\end{minipage}
\end{center}
\label{tab:pn150}
\end{table*}
\renewcommand{\baselinestretch}{1.5}

\renewcommand{\baselinestretch}{1.}
\begin{table*}
\begin{center}
\begin{minipage}{11.5cm}
\caption{Lexington 2000 optically thin planetary nebula (PN75) benchmark case results. }
\label{tab:pn75}
\begin{tabular}{lr|ccccccc}
\multicolumn{8}{c}{} \\
\hline
Line				& Med & GF	& HN	& DP	& PH	& RS	& \multicolumn{2}{c}{BE}\\
				&	&	&	&	&	&	& 3-D	& 1-D \\
\hline
H$_{\beta}$/10$^{34}$ erg/s	& 5.71	& 6.08	& 5.56	& 5.74	& 5.96	& 5.69	& 5.65	& 5.63 \\
H$_{\beta}$ 4861		& -	& 1.00	& 1.00	& 1.00	& 1.00 	& 1.00 	& 1.00  & 1.00 \\
He~{\sc i} 5876     		& 0.131	& 0.130	& 0.144	& 0.132	& 0.126	& 0.125	& 0.132 & 0.132 \\
He~{\sc ii} 4686		& 0.087	& 0.085	& 0.089	& 0.087	& 0.087	& -	& 0.093 & 0.094 \\
C~{\sc ii}$]$ 2325+		& 0.039	& 0.023	& 0.047	& 0.040	& 0.044	& 0.034	& 0.038 & 0.043 \\
C~{\sc ii} 1335			& 0.089	& 0.096	& 0.089	& 0.101	& 0.085	& -	& 0.086 & 0.085 \\
C~{\sc iii}$]$ 1907+1909 	& 0.790	& 0.584	& 0.96	& 0.882	& 0.602	& 1.00	& 0.698 & 0.709 \\
C~{\sc iv} 1549+		& 0.354	& 0.298	& 0.480	& 0.393	& 0.291	& 0.315	& 0.414 & 0.463 \\
$[$N~{\sc ii}$]$ 6584+6548	& 0.098	& 0.069	& 0.097	& 0.089	& 0.108	& 0.119	& 0.100 & 0.087 \\
$[$N~{\sc ii}$]$ 5755		& .0012	& -	& .0011	& .0012	& .0013	& .0020	& .0011 & .0010 \\
 N~{\sc iii}$]$ 1749+		& 0.043	& 0.029	& 0.059	& 0.056	& 0.038	& 0.048	& 0.038 & 0.039 \\
$[$N~{\sc iii}$]$ 57.3 $\mu$m	& 0.397	& 0.371	& 0.405	& 0.404	& 0.390	& 0.405	& 0.336 & 0.334 \\
N~{\sc iv}$]$ 1487+		& 0.018	& 0.019	& 0.024	& 0.020	& 0.012	& 0.011	& 0.017 & 0.020 \\
$[$O~{\sc ii}$]$ 3726+3729	& 0.262	& 0.178	& 0.262	& 0.266	& 0.262	& 0.311	& 0.234 & 0.205 \\
$[$O~{\sc iii}$]$ 5007+4959	& 11.35	& 10.1	& 13.2	& 11.7	& 10.1	& 11.8	& 11.0  & 11.1 \\
$[$O~{\sc iii}$]$ 4363		& 0.060	& 0.046	& 0.077	& 0.066	& 0.048	& 0.065	& 0.056 & 0.057 \\
$[$O~{\sc iii}$]$ 51.8 $\mu$m 	& 1.98	& 1.94	& 2.09	& 1.94	& 1.95	& 2.02	& 2.07  & 2.07 \\
$[$O~{\sc iii}$]$ 88.3 $\mu$m 	& 1.12	& 0.986	& 1.13	& 1.12	& 1.07	& 1.12	& 1.14  & 1.14 \\
$[$O~{\sc iv}$]$ 25.9 $\mu$m	& 0.814& 0.767	& 0.741	& 0.859	& 0.821	& 0.807	& 0.894 & 0.942 \\
O~{\sc iv}$]$ 1403+		& 0.013 & 0.009	& 0.015	& 0.014	& .093	& -	& 0.013 & 0.015 \\
$[$Ne~{\sc ii}$]$ 12.8 $\mu$m 	& 0.012 & 0.012	& 0.012	& 0.012	& 0.012	& 0.017	& 0.013 & 0.012 \\
$[$Ne~{\sc iii}$]$ 15.5 $\mu$m 	& 0.948	& 0.883	& 0.95	& 0.902	& 1.32	& 1.35	& 0.946 & 0.949 \\
$[$Ne~{\sc iii}$]$ 3869+3968	& 0.872	& 0.784	& 0.948	& 0.818	& 0.919	& 1.10	& 0.826 & 0.838 \\
$[$Ne~{\sc iv}$]$ 2423+		& 0.030	& 0.028	& 0.032	& 0.036	& 0.027	& 0.020	& 0.034 & 0.039 \\
Mg~{\sc ii} 2798+		& 0.102	& 0.086	& 0.14	& 0.111	& 0.071	& 0.093	& 0.114 & 0.106 \\
$[$Mg~{\sc iv}$]$ 4.49 $\mu$m 	& .0062 & .0021	& .006	& .0075	& .0065	& .0050	& .0068 & .0072 \\
$[$Si~{\sc ii}$]$ 34.8 $\mu$m	& 0.029	& 0.025	& 0.034	& 0.025	& 0.060	& 0.004	& 0.061 & 0.052 \\
Si~{\sc ii}$]$ 2335+		& .0057	& .0037	& .0078	& .0054	& -	& .0010	& .0062 & .0052 \\
Si~{\sc iii}$]$ 1892+		& 0.104	& 0.087	& 0.16	& 0.136	& 0.101	& 0.019	& 0.107 & 0.110 \\
Si~{\sc iv} 1397+		& 0.017	& 0.017	& 0.023	& 0.018	& 0.013	& 0.023	& 0.016 & 0.018 \\
$[$S~{\sc ii}$]$ 6716+6731	& .0020	& 0.023	& 0.036	& 0.029	& 0.013	& 0.016	& 0.017 & 0.013 \\
$[$S~{\sc ii}$]$ 4069+4076	& .0017	& .0022	& .0034	& .0030	& .0013	& .0010	& .0012 & .0010 \\
$[$S~{\sc iii}$]$ 18.7 $\mu$m	& 0.486	& 0.619	& 0.715	& 0.631	& 0.316	& 0.357	& 0.285 & 0.266 \\
$[$S~{\sc iii}$]$ 33.6 $\mu$m	& 0.533 & 0.702	& 0.768	& 0.684	& 0.339	& 0.383	& 0.306 & 0.285 \\
$[$S~{\sc iii}$]$ 9532+9069	& 1.20	& 1.31	& 1.51	& 1.33	& 0.915	& 1.09	& 0.831 & 0.777 \\
$[$S~{\sc iv}$]$ 10.5 $\mu$m	& 1.94  & 1.71	& 1.57	& 1.72	& 2.17	& 2.33	& 2.79  & 2.87 \\
10$^3$\,$\times$\,$\Delta$(BC 3645)/\AA & 4.35 &  4.25 & - & 4.25	& 4.35	& 4.90	& 4.54 & 4.56 \\ 
$T_{\rm inner}$/ K   		& 14300 & 14450	& 14640	& 14680	& 14150	& 13620	& 14100 & 14990 \\ 
$<$T$[$N$_{\rm p}$N$_{\rm e}]>$/K 	& 10425 & 9885	& 11260	& 10510	& 10340	& 10510	& 10220 & 10263 \\
$R_{\rm out}$/10$^{17}$cm		& -	& 7.50	& 7.50	& 7.50	& 7.50	& 7.50	& 7.50 & 7.50 \\
$<$He$^+>/<$H$^+>$		       	& 0.913 & 0.912 & 0.92 & 0.914	& 0.920	& 0.913	& 0.911 & 0.908 \\
$\tau$(1Ryd)			& 1.47  &  1.35	& 1.64	& 1.61	& 1.47	& -	& 1.15 & 1.29 \\
\hline
\end{tabular}
\small{GF: G. Ferland's {\it Cloudy}; PH: J.P Harrington code; DP: D. P\'equignot's {\it Nebu}; HN: H. Netzer's {\it Ion}; RS: R. Sutherland's {\it Mappings}; BE: B. Ercolano's {Mocassin}.}
\end{minipage}
\end{center}
\end{table*}
\renewcommand{\baselinestretch}{1.5}

The Lexington/Meudon Standard H~{\sc ii}~region model (HII40) was the first benchmark to be run and some very preliminary results have already been presented, at the November 2000 Lexington meeting \citep{ercolano01, pequignot01}. However, those results were produced when Mocassin was still under development and should therefore only be considered as a {\it snapshot} of the code at that particular stage. The code has evolved considerably 
since the November 2000 Lexington meeting and the newer results are presented in this section (see Table~\ref{tab:hii40}). 

Table~\ref{tab:anMon} shows the results of a comparison between the line fluxes obtained by Mocassin using the formal solution method and those obtained using the Monte Carlo method (see Section~\ref{sub:comparison}) for some of the more significant lines in the benchmark cases. It is clear that the results shown agree well; however, as expected, larger discrepancies were found for the weaker lines, whose lower numbers of energy packets yield lower accuracy statistics. 

\subsubsection{The HII40 benchmark}

Mocassin scored eight {\it if}'s~$>$~1.01 for the HII40 benchmark model (Table~\ref{tab:ifs}); only three of these, however, had values greater than 1.3. Amongst these, {\it if}'s~$>$~1.10 are obtained for the $[$O~{\sc iii}$]$ 5007+4959 ({\it if} = 1.18) and for $[$O~{\sc iii}$]$ 4363 ({\it if} = 1.45); the ratio of these lines is often used as a  temperature diagnostic \citep[see, for example,][pages 119-125]{osterbrock89}. Mocassin predicts $\frac{j_{\lambda4959}+j_{\lambda5007}}{j_{\lambda4363}}$~=~745.4, this value is higher than the value obtained by the other codes, in fact median value obtained for the ratio of these line fluxes by the other codes is equal to 589.2. This is fully consistent with Mocassin predicting a slightly lower temperature ({\it if} = 1.027) for this benchmark than do the other codes. 

Finally, the number of median values obtained for this benchmark case is ten, which compares very well with the other codes' median scores, ranging from three to ten (see Table~\ref{tab:medians}).

\begin{figure*}
\begin{center}
\begin{minipage}[t]{7.3cm} 
\psfig{file=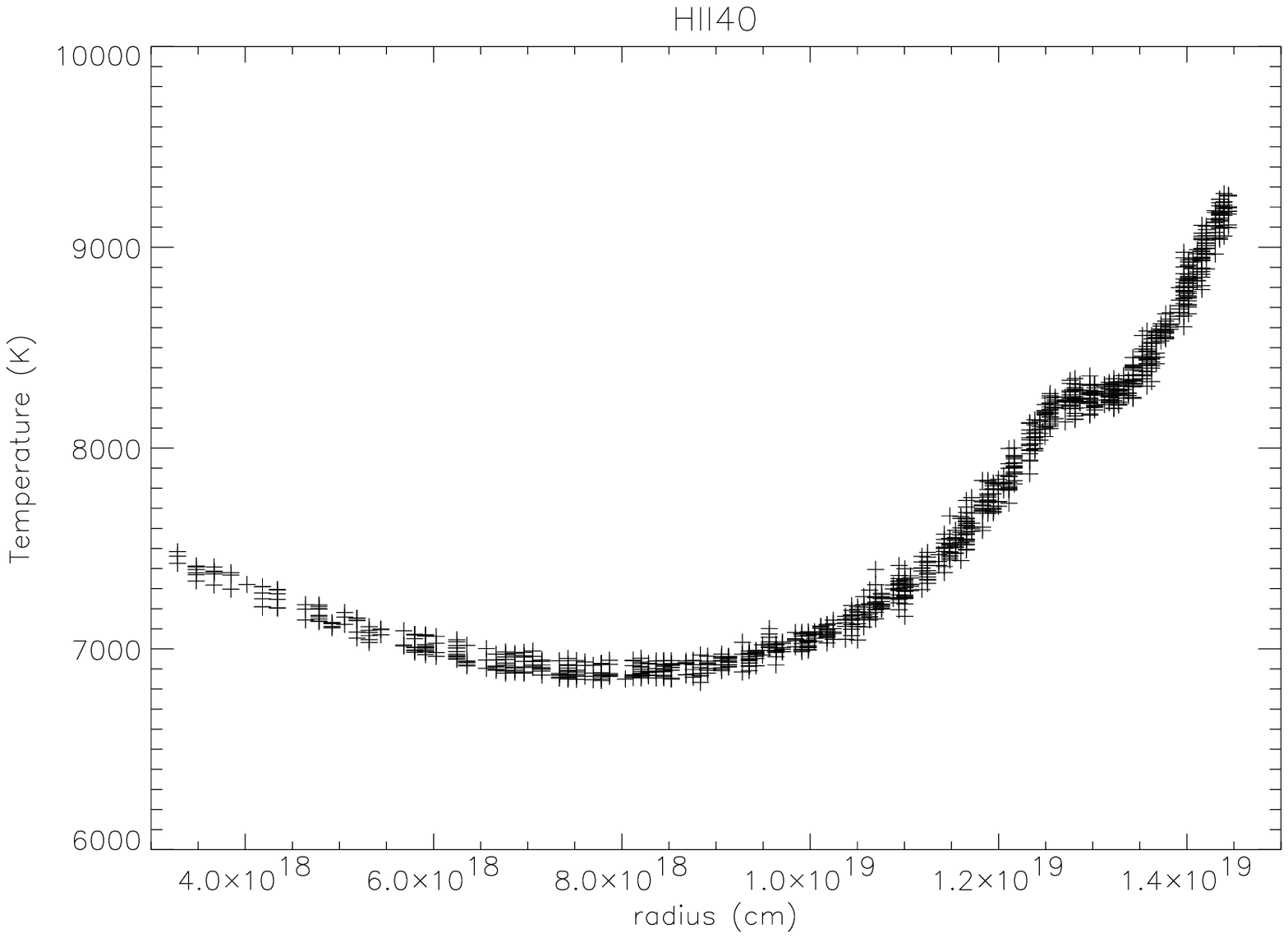, height=73mm, width=70mm}
\end{minipage}
\begin{minipage}[t]{7.3cm}
\psfig{file=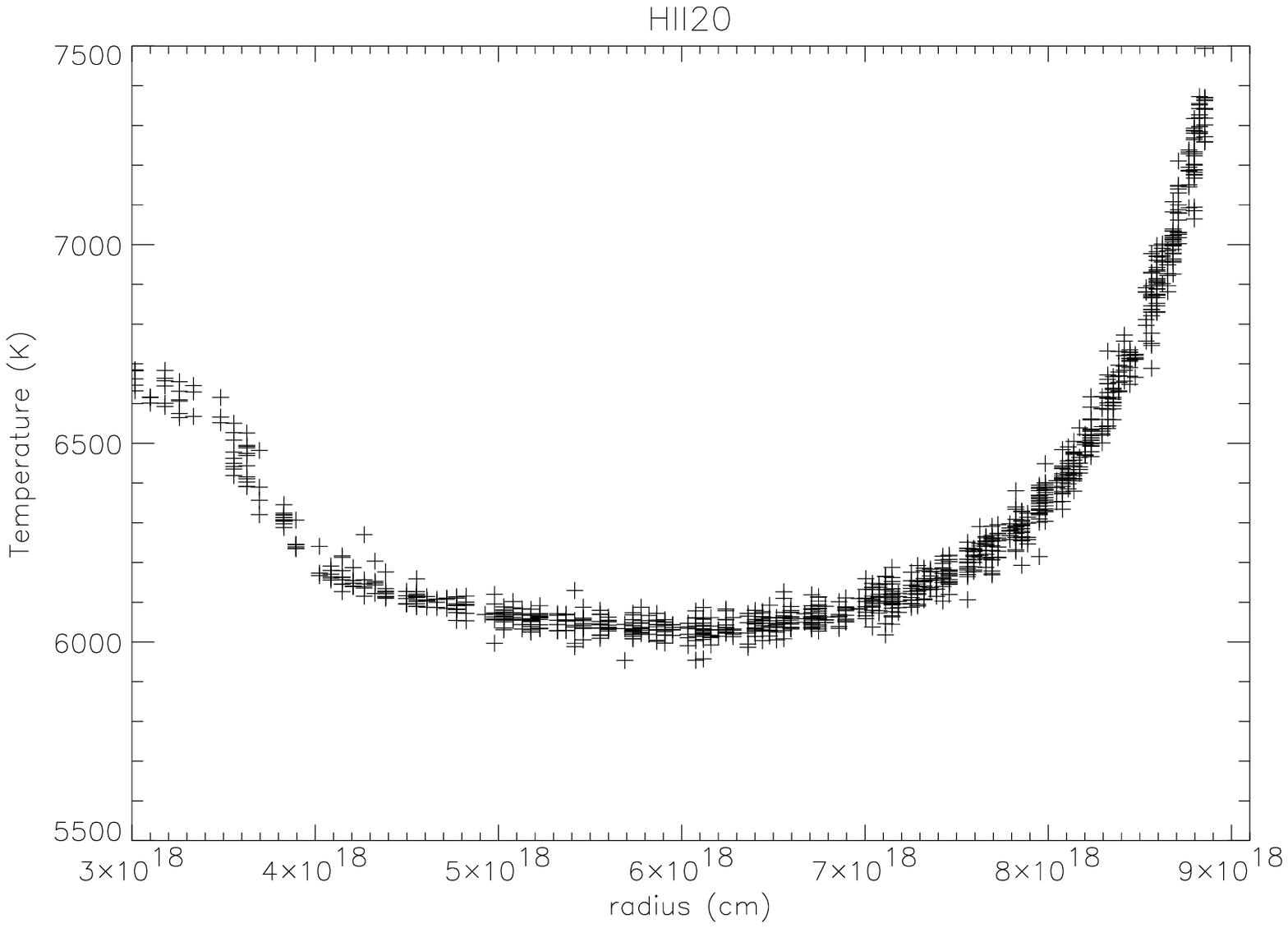, height=73mm, width=70mm}
\end{minipage}
\begin{minipage}[t]{7.3cm} 
\psfig{file=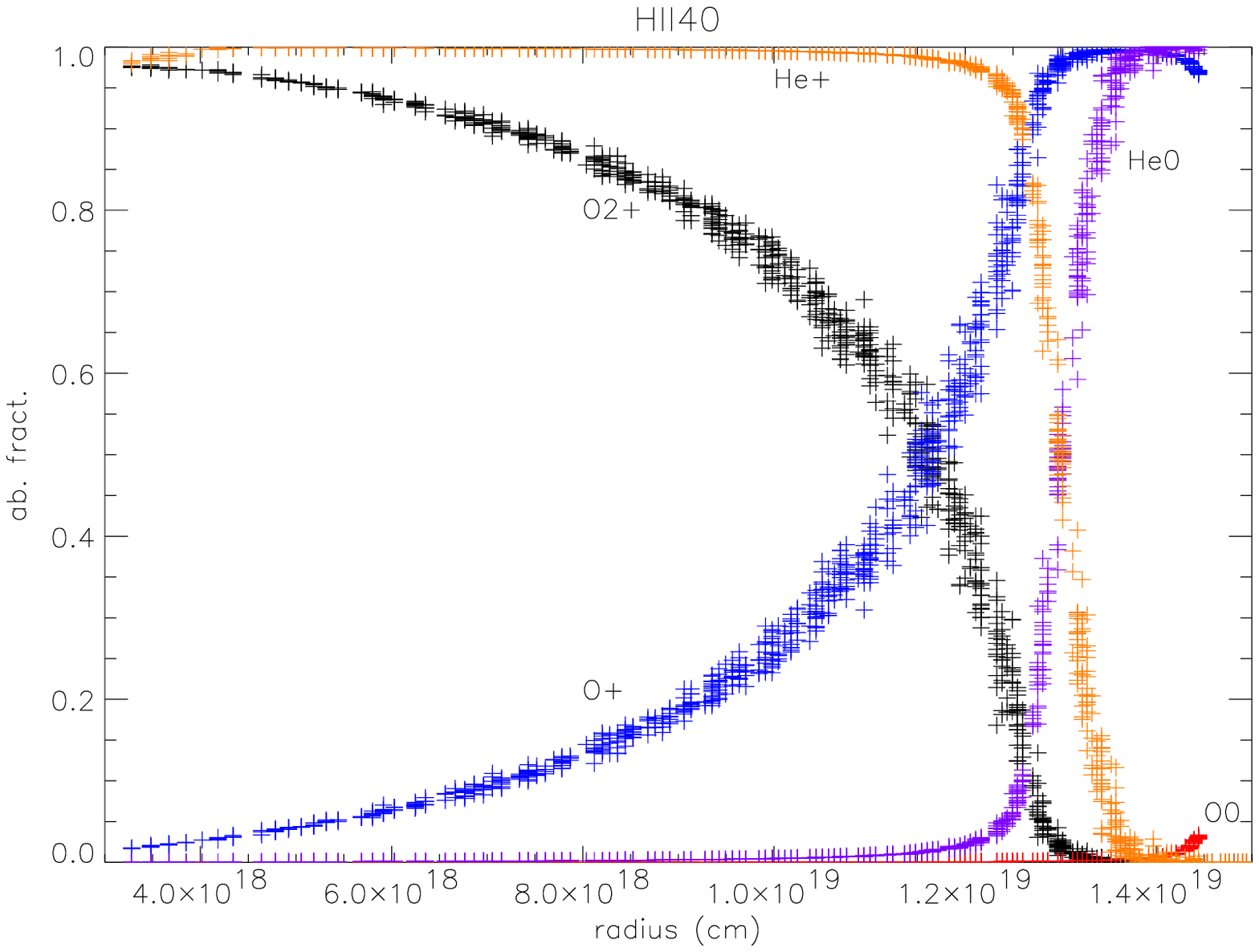, height=73mm, width=70mm}
\end{minipage}
\begin{minipage}[t]{7.3cm} 
\psfig{file=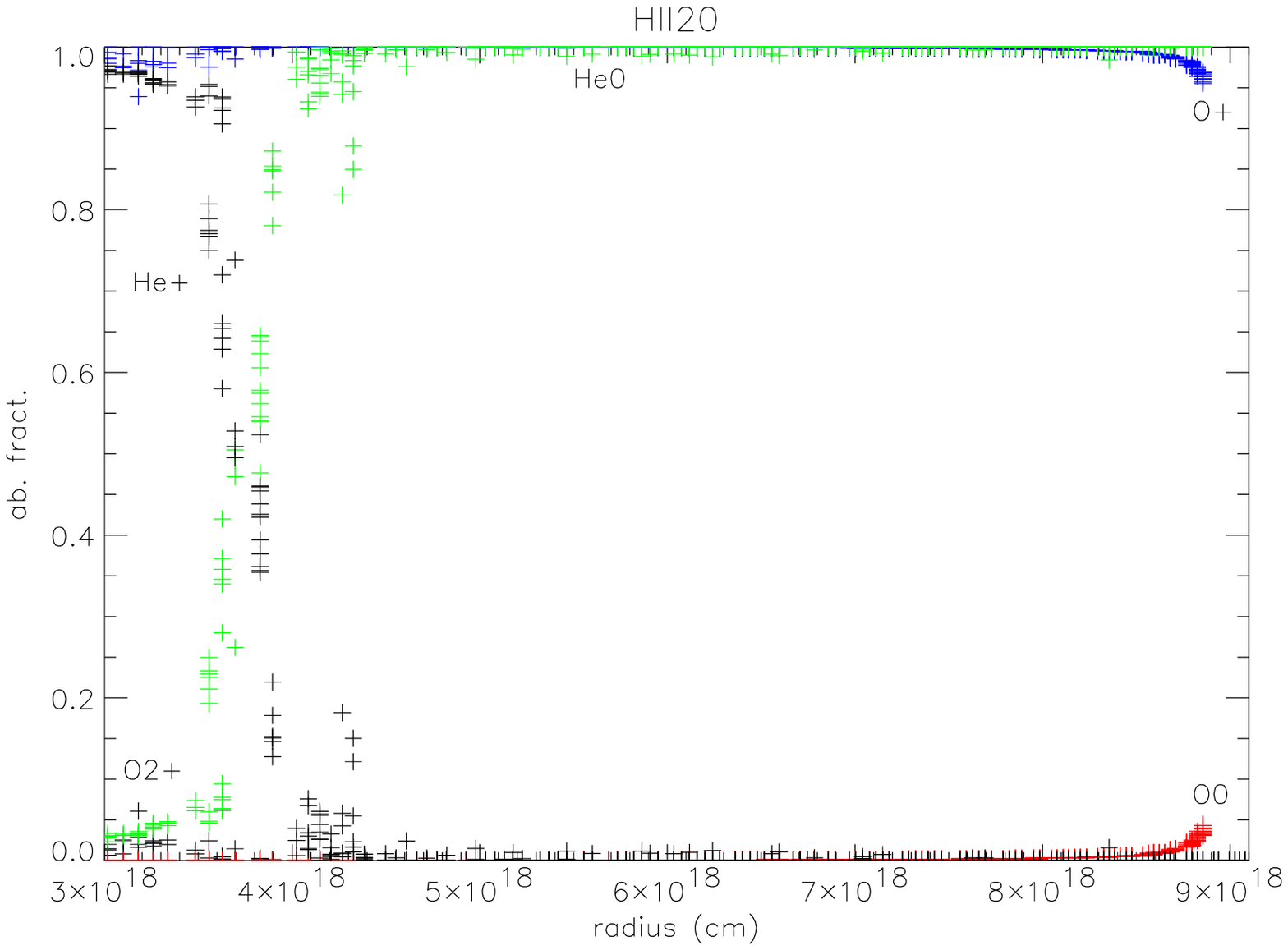, height=73mm, width=70mm}
\end{minipage}
\begin{minipage}[t]{7.3cm}
\psfig{file=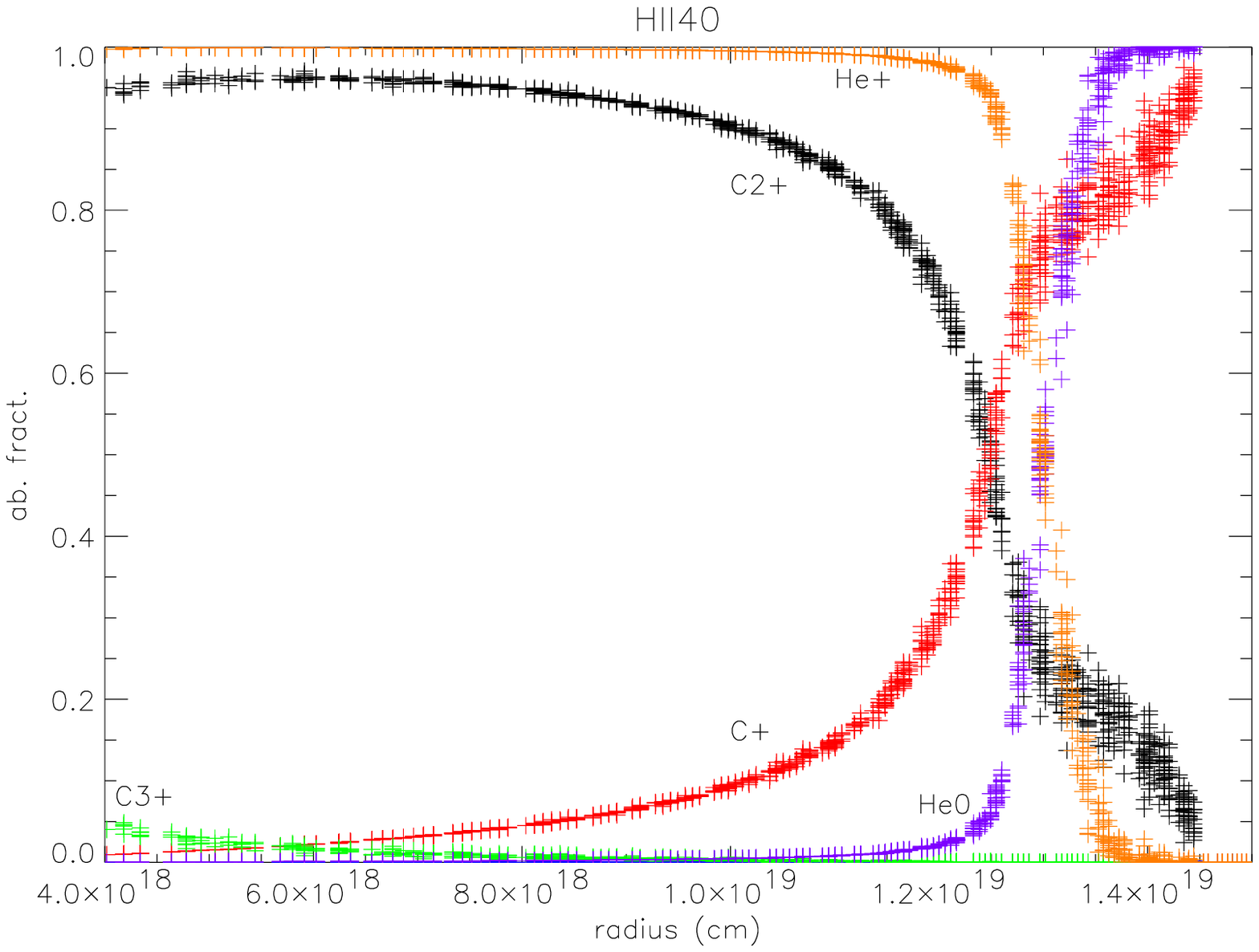, height=73mm, width=70mm}
\end{minipage}
\begin{minipage}[t]{7.3cm}
\psfig{file=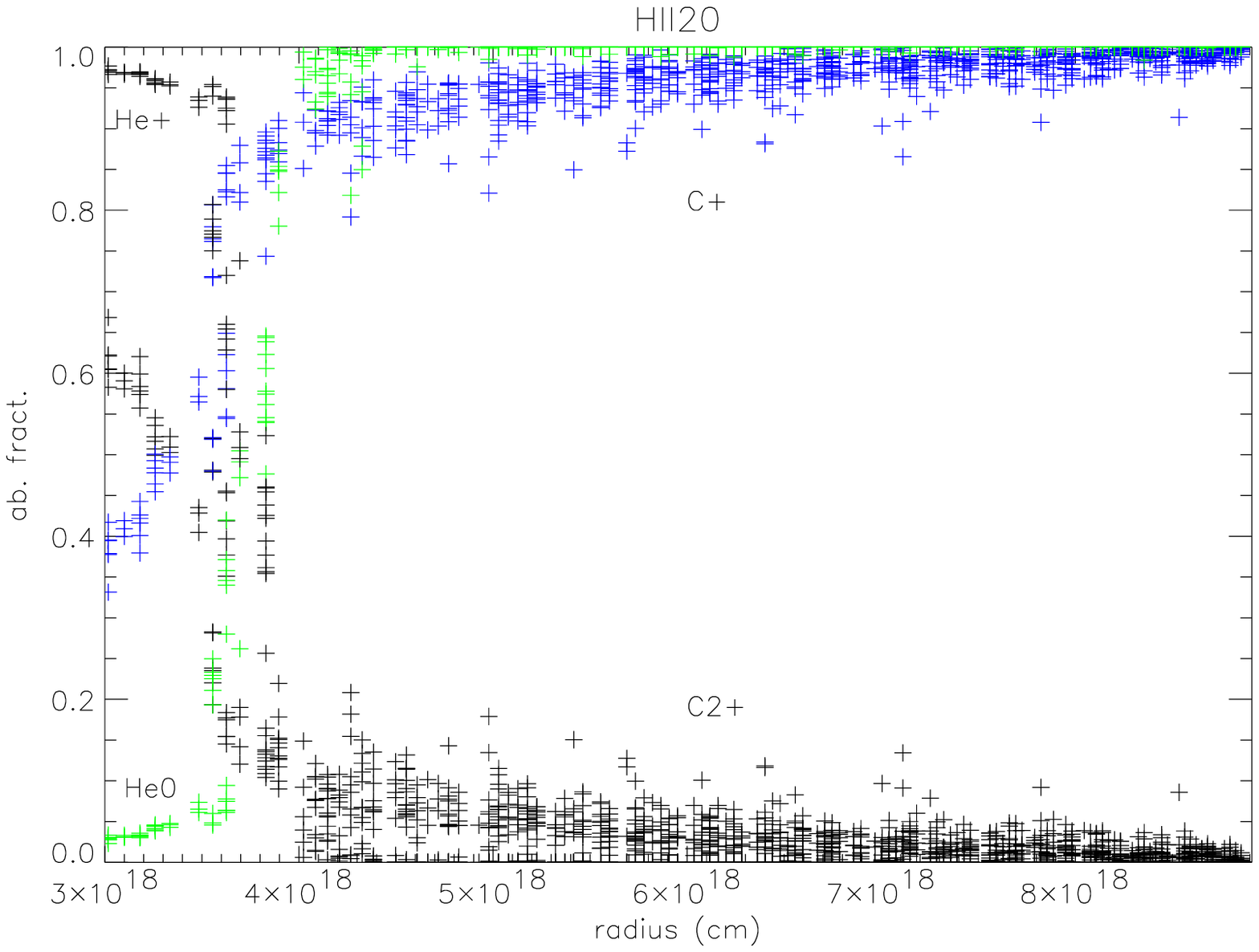, height=73mm, width=70mm}
\end{minipage}
\caption[Fractional ionic abundances of O and C for the H~{\sc ii} region benchmark cases.]{Electron temperature (top panels) and the fractional ionic abundances of oxygen (middle panels) and carbon (bottom panels), as a function of nebular radius, for the H~{\sc ii} region benchmark cases HII40 (left-hand panels) and HII20 (right-hand panels).}
\label{fig:xhii}
\end{center}
\end{figure*}

\subsubsection{The HII20 benchmark}

Mocassin scored seven {\it if}'s for the low excitation H~{\sc ii} region (HII20) benchmark model. None of these, however, has a value greater than 1.3. As in the HII40 nechmark case, the mean temperature, weighted by $N_{\rm p}N_{\rm e}$, predicted by Mocassin for this model is also slightly lower ({\it if} = 1.034) than the other models' predictions. 

Five median values were obtained by Mocassin for this benchmark case, while the other codes scored between three and eleven (see Table~\ref{tab:medians}). 

\subsubsection{The PN150 benchmark}

\begin{figure*}
\begin{center}
\begin{minipage}[t]{7.3cm} 
\psfig{file=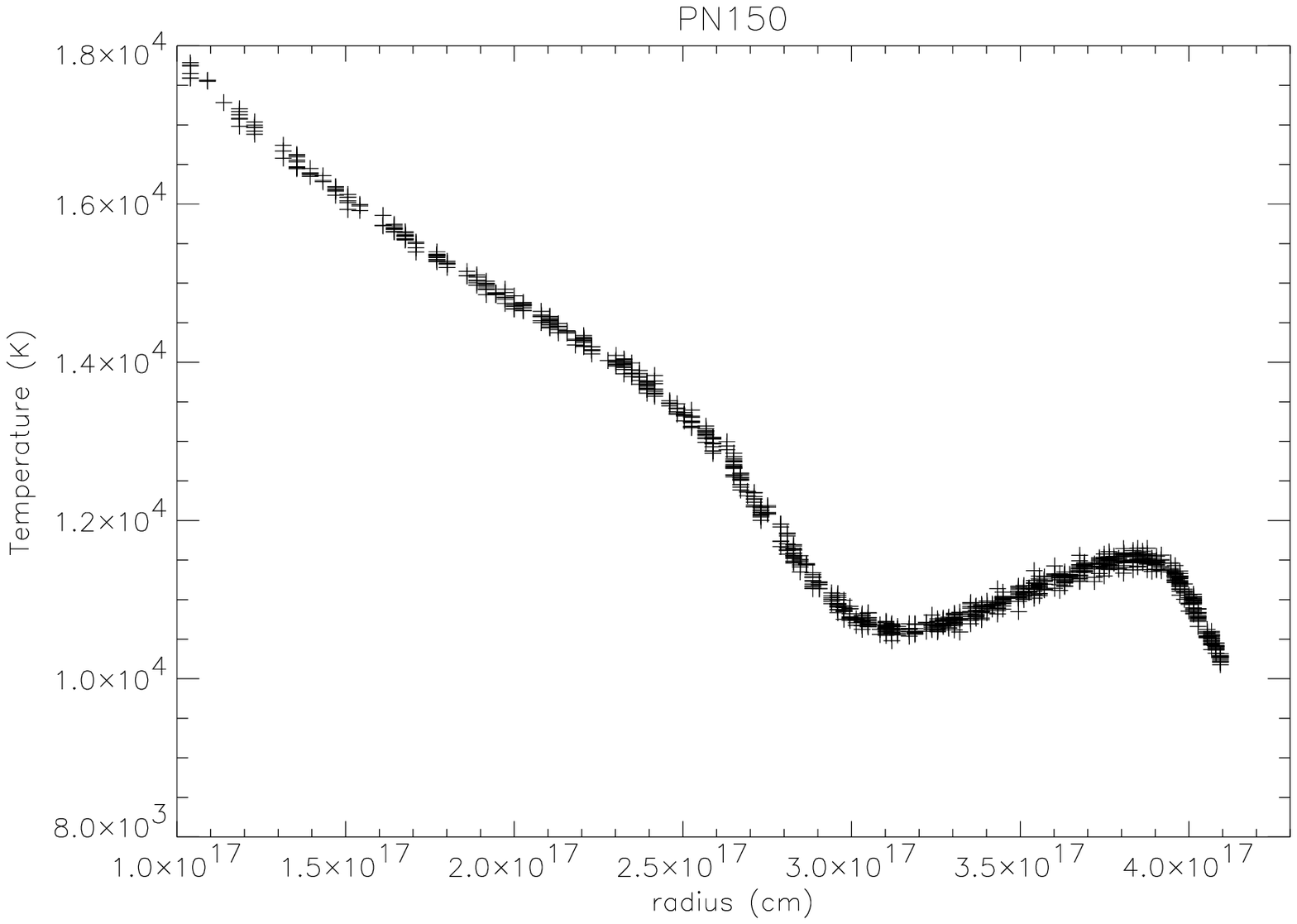, height=73mm, width=73mm}
\end{minipage}
\begin{minipage}[t]{7.3cm}
\psfig{file=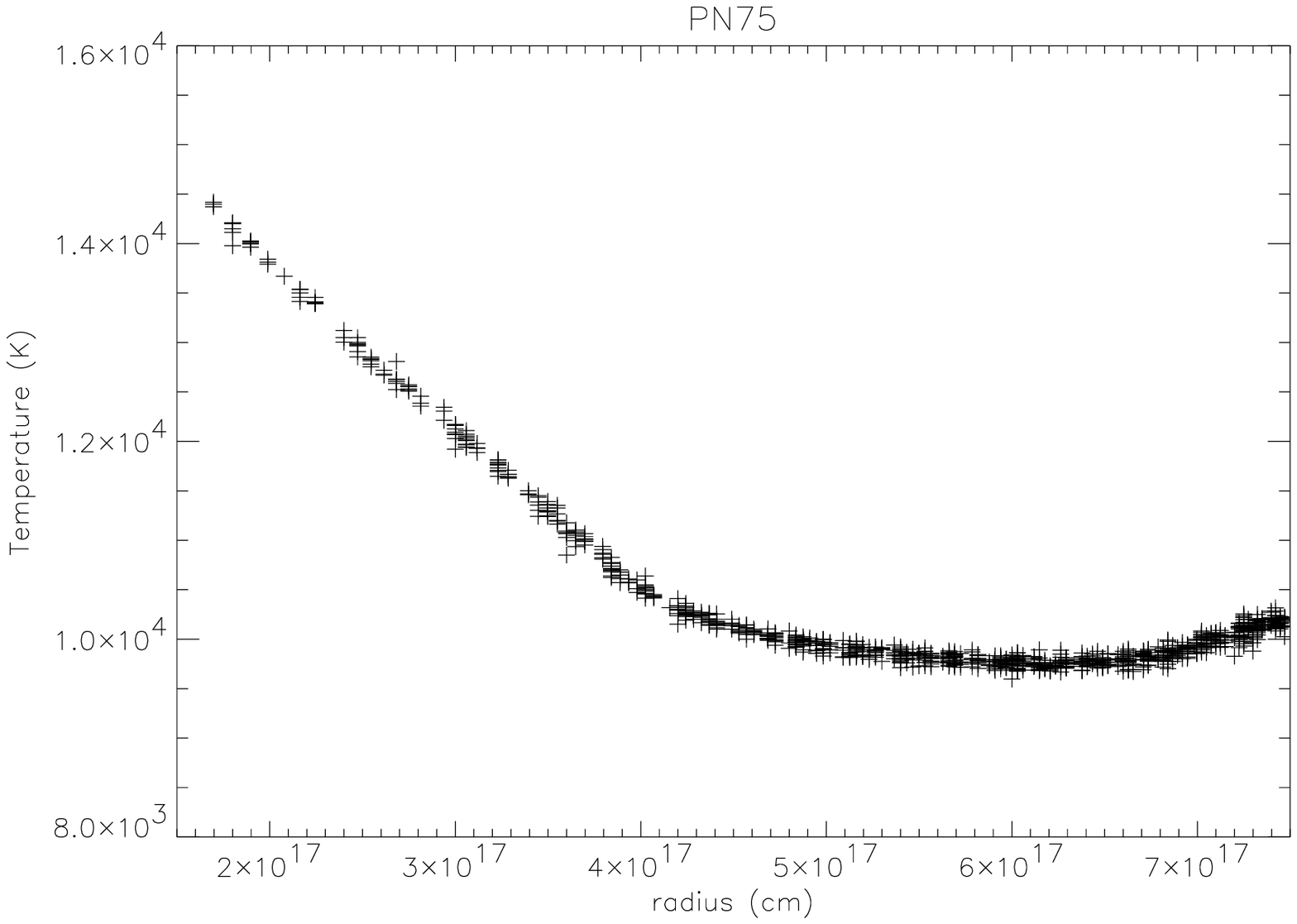, height=73mm, width=73mm}
\end{minipage}
\begin{minipage}[t]{7.3cm} 
\psfig{file=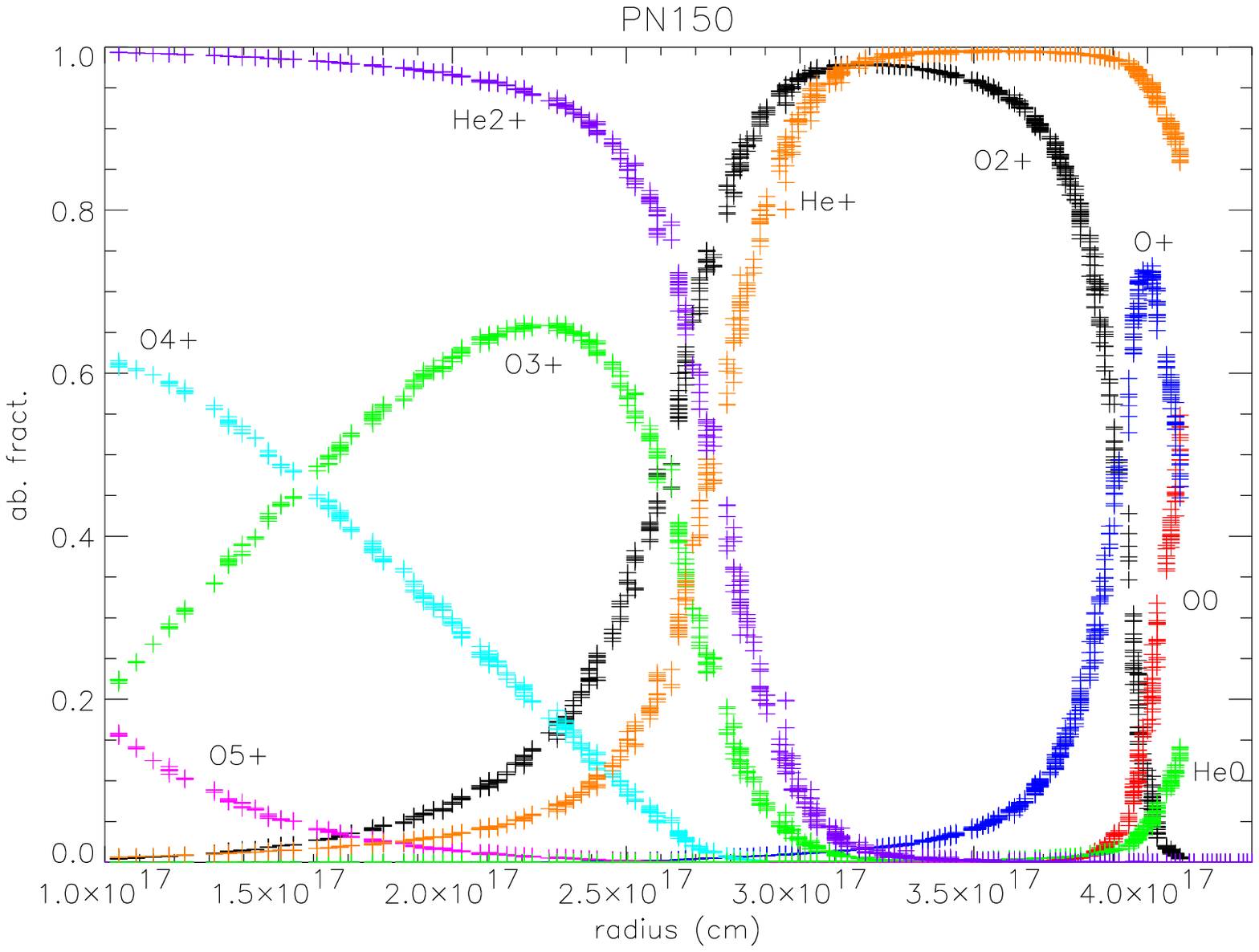, height=73mm, width=73mm}
\end{minipage}
\begin{minipage}[t]{7.3cm} 
\psfig{file=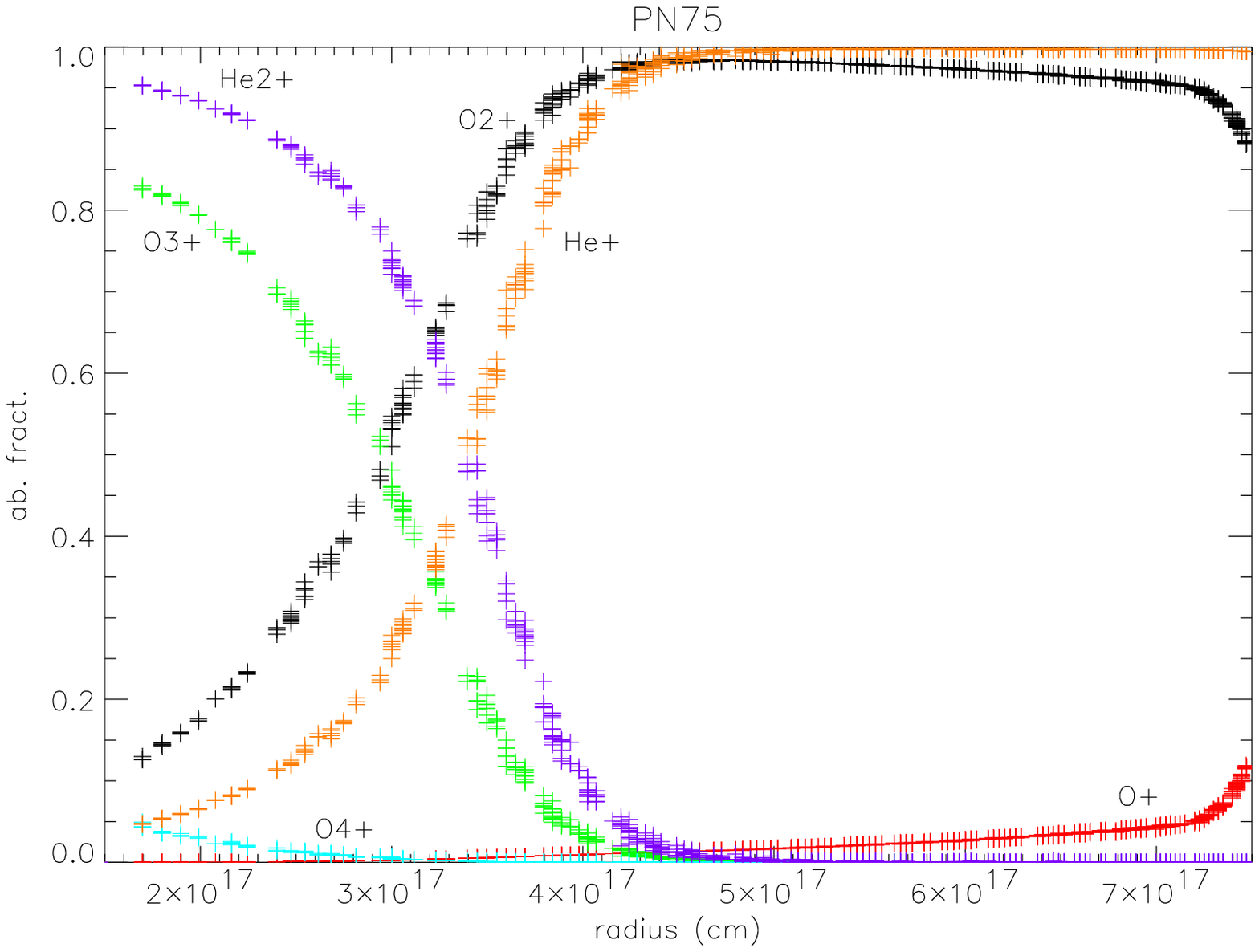, height=73mm, width=73mm}
\end{minipage}
\begin{minipage}[t]{7.3cm}
\psfig{file=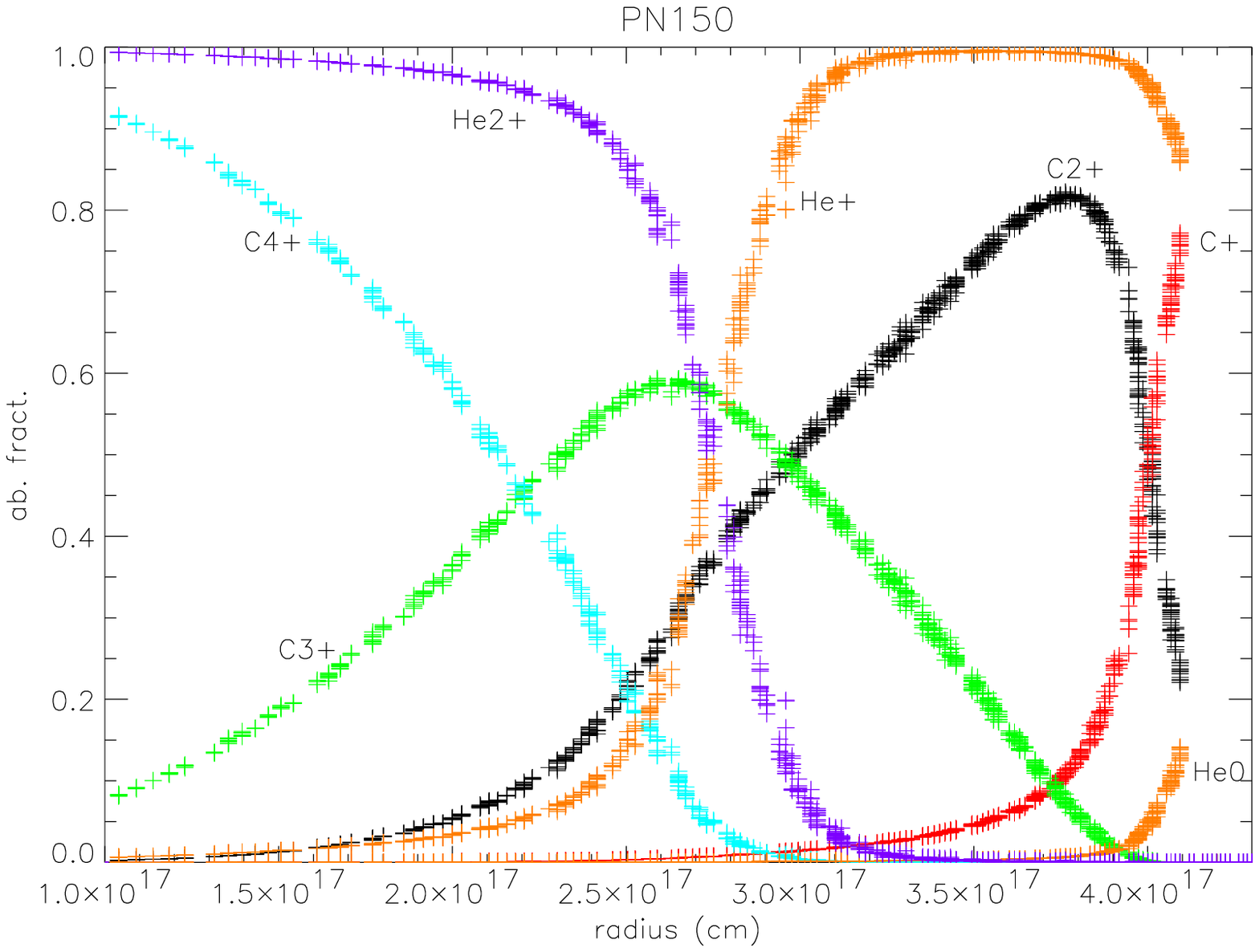, height=73mm, width=73mm}
\end{minipage}
\begin{minipage}[t]{7.3cm}
\psfig{file=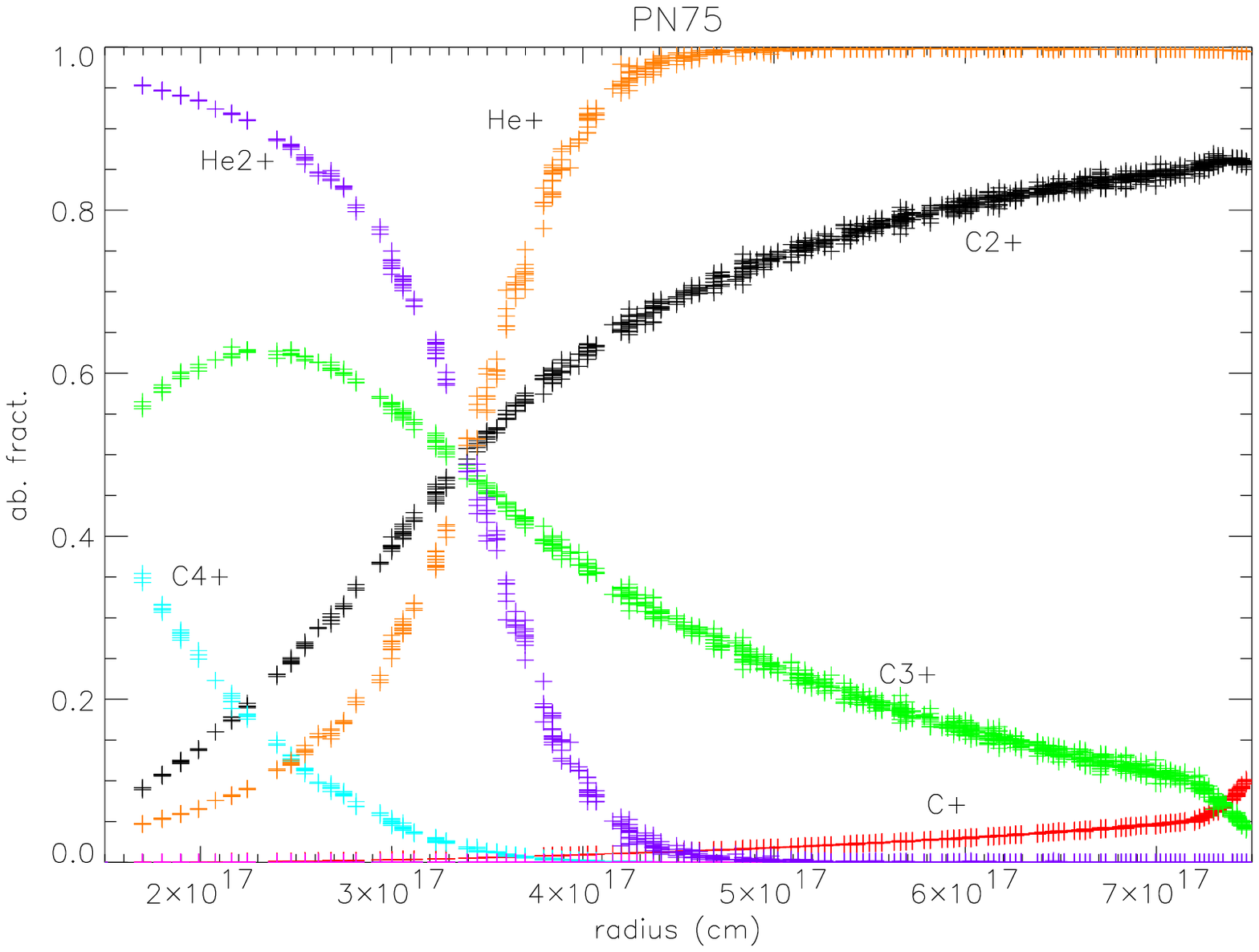, height=73mm, width=73mm}
\end{minipage}
\caption[Fractional ionic abundances of O and C for the PN benchmark cases.]{Electron tempearture (top panels) and the fractional ionic abundances of oxygen (middle panels) and carbon (bottom panels), as a function of nebular radius, for the planetary nebula benchmark cases PN150 (left-hand panels) and PN75 (right-hand panels).}
\label{fig:xpn}
\end{center}
\end{figure*}

The optically thick high-excitation planetary nebula (PN150) is the most demanding of the benchmark cases in terms of physical processes and atomic data required. Mocassin's score for this model was very good (Table~\ref{tab:ifs}), obtaining only six {\it if}'s, with none of those being higher than 1.3 and only one slightly higher than 1.1 (C~{\sc ii}~$\lambda1335$, {\it if}~=~1.13). As has already been discussed by \citet{pequignot01}, there seems to be a dichotomy between the GF, HN and DP models (and, now, also the BE model) on the one hand, and the TK, PH and RS models on the other. The former group obtained very few {\it if}'s largest than 1.1, indicative of a tighter agreement. This coherence can probably be attributed to a more recent updating of atomic data and a more careful treatment of the diffuse radiation field transfer. These four codes also obtained a larger H$\beta$ flux for this model, which can probably be ascribed to secondary photons from heavy ions. PH is the only classical code here with a fully iterative spherically symmetric radiative transfer treatment (since RR only computed H~{\sc ii} regions); this could also be the reason for the relatively larger number of {\it if}'s scored by the PH code for this model. 

The score for median values obtained by Mocassin for the PN150 optically thick planetary nebula is extremely good, obtaining the highest value of fifteen medians, above the other codes which obtained between two and thirteen (see Table~\ref{tab:medians}).

\subsubsection{The PN75 benchmark}

The optically thin planetary nebula (PN75) benchmark model is not a radiation bounded case, but a matter bounded one and, in fact, the outer radius is given as an input parameter to all codes and fixed at $7.5\times10^{19}$\,cm. For this reason, for this particular model there is not a straightforward {\it conservation law} for the absolute flux of H$\beta$. This can be used to explain, in part at least,  the relatively poor scores of the GF code for low {\it if}'s (Table~\ref{tab:ifs}), since, for one reason or another, its predicted H$\beta$ flux deviated somewhat from the median value, thus shifting all the other line intensities, given in H$\beta$ units. The PH code also obtained an H$\beta$ flux which deviated from the median value; in this case, however, the number of total {\it if}'s stayed low (=5) and no {\it if}~$>$~1.30 was obtained. Mocassin, however, obtained a low number of {\it if}'s for this relatively difficult case, scoring nine {\it if}'s in total, with none of those having a value greater than 1.30. 

Mocassin obtained a score of thirteen median values for this benchmark case, which compares well with the scores obtained by the other codes for this benchmark, ranging from eight to eighteen median values.

\section{Discussion}
\label{sec:discussion}

The overall performance of Mocassin for the four benchmarks was very satisfactory, as shown by Table~\ref{tab:ifs}. The results obtained from the one-dimensional mode of Mocassin are, in general, in very good agreement with those obtained using the fully three-dimensional Mocassin models. One noticable difference, common to all four benchmarks, is that the kinetic temperature at the illuminated inner edge of the nebula, T$_{inner}$, is higher for the one-dimensional Mocassin results and closer to the values obtained by the other one-dimensional codes included in the comparison. This is an obvious effect caused by the coarseness of the three-dimensional grid: since all the physical properties of the gas are constant within each volume element, then the electron temperature of a given cell will be mainly representative of the kinetic temperature at its centre. From this, it naturally follows that the coarser the grid is, and the larger the cells, then the  further the kinetic temperature at the centres of the cells adjacent to the inner radius will be from the true value at the inner radius.

\renewcommand{\baselinestretch}{1.2}
\begin{table}
\begin{center}
\caption{Summary of isolation factors, $if$'s, for the benchmark cases}
\begin{tabular}{lcccccccc}
\multicolumn{9}{c}{} \\
\hline
Case 				& GF	& HN	& DP	& TK	& PH	& RS	& RR	& BE\\
\hline
HII40				&	&	&	&	&	&	&	&   \\
$>$ 1.01			& 8	& 5	& 1	& 17	& 2	& 4	& 7	& 8 \\
$>$ 1.03			& 5	& 3	& 0	& 15	& 1	& 3	& 5	& 6 \\
$>$ 1.10			& 3	& 0	& 0	& 8	& 0	& 2	& 1	& 5 \\
$>$ 1.30			& 0	& 0	& 0	& 5	& 0	& 0	& 0	& 3 \\
$>$ 2.00			& 0	& 0 	& 0	& 2	& 0	& 0	& 0	& 0 \\
No pred.			& 0	& 1	& 0	& 5	& 0	& 0	& 5	& 0 \\
\hline
HII20				& 	&	&	&	&	&	&	&   \\
$>$ 1.01			& 3	& 2	& 2	& 12	& 4	& 7	& 4	& 7 \\
$>$ 1.03			& 2	& 1	& 2	& 10	& 3	& 6	& 4	& 6 \\
$>$ 1.10			& 0	& 0	& 0	& 8	& 0	& 2	& 3	& 3 \\
$>$ 1.30			& 0	& 0	& 0	& 5	& 0	& 2	& 0	& 0 \\
$>$ 2.00			& 0	& 0	& 0	& 2	& 0	& 0	& 0	& 0 \\
No pred.			& 0	& 2	& 0	& 1	& 0	& 0	& 3	& 0 \\
\hline
PN150				& 	&	&	&	&	&	&	&   \\
$>$ 1.01			& 4	& 8	& 2	& 27	& 15	& 23	& -	& 6 \\
$>$ 1.03			& 4	& 6	& 2	& 26	& 13	& 19	& -	& 5 \\
$>$ 1.10			& 1	& 2	& 0	& 22	& 7	& 16	& -	& 1 \\
$>$ 1.30			& 0	& 0	& 0	& 17	& 6	& 7	& -	& 0 \\
$>$ 2.00			& 0	& 0	& 0	& 7	& 2	& 2	& -	& 0 \\
No pred.			& 0	& 0	& 0	& 3	& 1	& 6	& - 	& 0 \\
\hline
PN75				& 	&	&	&	&	&	&	&   \\
$>$ 1.01			& 14	& 20	& 4	& -	& 5	& 14	& -	& 9 \\
$>$ 1.03			& 11	& 18	& 4	& -	& 4	& 13	& -	& 8 \\
$>$ 1.10			& 5	& 10	& 2	& -	& 4	& 10	& -	& 6 \\
$>$ 1.30			& 4	& 1	& 0	& -	& 0	& 3	& -	& 0 \\
$>$ 2.00			& 1	& 0	& 0	& -	& 0	& 3	& -	& 0 \\
No pred.				& 1	& 1	& 0	& -	& 1	& 4	& -	& 0 \\
\hline
\end{tabular}
\label{tab:ifs}
\end{center}
\end{table}
\renewcommand{\baselinestretch}{1.5}

\begin{table}
\begin{center}
\caption{Summary of median values for the benchmark cases}
\begin{tabular}{lr|cccccccc}
\multicolumn{10}{c}{} \\
\hline
Case		& Total & GF	& HN	& DP	& TK	& PH	& RS	& RR	& BE \\
\hline	
HII40		& 31	& 8	& 8	& 10	& 3	& 9	& 9	& 5	& 10  \\
HII20		& 24	& 4	& 7	& 7	& 3	& 11	& 6	& 4	& 5  \\
\hline
\multicolumn{2}{l}{Subtot HII}  & 12	& 15	& 17	& 6	& 20	& 15	& 9	& 15 \\
\hline
PN150		& 49	& 9 	& 11	& 12	& 2	& 13	& 4	& -	& 15 \\
PN75		& 40	& 10 	& 8	& 19	& -	& 16	& 13	& -	& 13 \\
\hline
\multicolumn{2}{l}{Subtot PN}	& 20	& 19	& 31	& (2)	& 29	& 17	& -	& 28 \\
\hline
\end{tabular}
\label{tab:medians}
\end{center}
\end{table}

The electron temperatures, $<T[N_{\rm p}N_{\rm e}]>$ and $T_{\rm inner}$, predicted by Mocassin for the Lexington benchmark models tend, in particular in the H~{\sc ii} region cases, towards the lower limit of the scatter. In the case of $T_{\rm inner}$, this seems to be a characteristic of all codes using an exact treatment for the radiative transfer. As noted by \citet{pequignot01}, the kinetic temperatures calculated by codes with exact transfer tend to be lower in the innermost layers of the nebula, as the ionizing radiation field there is softer. Only two codes in the Lexington benchmarks treated the radiative transfer exactly, namely  Rubin's {\it Nebula} (RR) and the Harrington code (PH) and, in fact, Mocassin's results for the kinetic temperatures generally agree better with those two codes' predictions. For the standard H~{\sc ii} region benchmark (HII40), Mocassin's kinetic temperature at the inner edge of the nebula, T$_{inner}$, agrees extremely well with the predictions of the RR and PH codes. Similar results are obtained for the low excitation H~{\sc ii} region benchmark (HII20), where, again, Mocassin's T$_{inner}$ agrees with the results of PH and RR. In both H~{\sc ii} regions benchmark cases, however, Mocassin predicted a  value which was about 250~K lower than the median for $<T[N_{\rm p}N_{\rm e}]>$, obtaining an {\it if}~=~1.027 for the HII40 case and {\it if}~=~1.034 for the HII20 case. The cause of this small discrepancy is not clear to us. 

Unfortunately, R. Rubin's code, {\it Nebula}, was not designed to treat planetary nebulae and, therefore, the only exact one-dimensional radiative transfer code available for the optically thick planetary nebula (PN150) and the optically thin planetary nebula (PN75) benchmarks is the Harrington code (PH). For PN150, Mocassin's T$_{inner}$ is in reasonable agreement with PH's prediction, particularly if the prediction from the one-dimensional Mocassin run is considered, since, as discussed earlier, this represents a measurement of the temperature taken closer to the inner edge of the nebula. The Mocassin result for $<T[N_{\rm p}N_{\rm e}]>$ is within the scatter and, in particular, BE and PH agree very well for this observable. Note that only HN and TK obtain higher temperatures for this model; moreover, the TK computation was carried out with a new code, still under development, primarily designed for X-ray studies. That code could not treat the diffuse radiation field, leading to problems for the hard radiation field cases, such as PN150. Finally, for the PN75 benchmark planetary nebula, Mocassin's T$_{inner}$ is within the scatter (the prediction from the one-dimensional model is actually at the higher limit of it) and in reasonable agreement with PH's prediction; the result for $<T[N_{\rm p}N_{\rm e}]>$ is also within the scatter and is in very good agreement with the prediction of the PH code. Once again, only HN predicts a higher value for this quantity, while TK's results for this model are not available.

The models presented in this chapter were all run using a 13$\times$13$\times$13 grid and, since they are all spherically symmetric, the ionizing source was placed in a corner of the grid. The number of energy packets used to sample the grids and bring them to convergence varied from three to five million. As has already been discussed, the accuracy of the results depends both on the spatial sampling (i.e. the number of grid cells) and on the number of energy packets used. It is clear, however, that the latter also depends on the number of points to be sampled, so if, for example, in a given simulation the number of grid cells is increased from $n_x$$\times$$n_y$$\times$$n_z$ to $n'_x$$\times$$n'_y$$\times$$n'_z$, then the number of energy packets used must also be increased from $N_{\rm packets}$ to $N'_{\rm packets}\,=\,\frac{n'_x\,\times\,n'_y\,\times\,n'_z}{n_x\,\times\,n_y\,\times\,n_z}\,\cdot\,N_{\rm packets}$. However for these relatively simple cases the three-dimensional grid specified above was found to be sufficient to produce acceptable results. In fact, since the benchmark models are spherically symmetric then, although the number of sampling points along each orthogonal axis is only 13, this is the equivalent of a one-dimensional code with 273 radial points, which is the number of different values of $r$ given by all the (x,y,z) combinations. This is clearly demonstated in figures~\ref{fig:xhii} to~\ref{fig:xpn}, where the number of data points and the spacing between them shows that the spatial sampling is indeed appropriate. The plots also show that the number of energy packets used in the simulations was sufficient, since the scatter of the ordinate values for a given $r$, which is essentially a measure of the error bars, is very small. The largest scatter was obtained in the plots for the HII20 benchmark (Figure~\ref{fig:xhii}); this is a very soft ionizing radiation field case and a larger number of energy packets is probably required in order to reduce the scatter shown and increase the accuracy of the results. For the purpose of this benchmark exercise, however, the accuracy achieved for HII20 is sufficient to produce satisfactory results. 

\section{Conclusions}

A fully three-dimensional photoionization code, Mocassin, has been developed for the modelling of photoionised nebulae, using Monte Carlo techniques. The stellar and diffuse radiation fields are treated self-consistently; moreover, Mocassin is completely independent of the assumed nebular geometry and is therefore ideal for the study of aspherical and/or inhomogeneous nebulae, or nebulae having one or more exciting stars at non-central locations.

The code has been successfully benchmarked against established one-dimensional photoinization codes for standard spherically symmetric model nebulae \citep[see][]{pequignot86, ferland95, pequignot01}. 

Mocassin is now ready for the application to real astronomical nebulae and it should provide an important tool for the construction of realistic nebular models. A companion paper \citep[][Paper~{\sc ii}]{ercolanoII} will present detailed results from the modelling of the non-spherically symmetric PN~NGC~3918. Resources permitting, it is intended to make the Mocassin source code publicly available in the near future.

\vspace{7mm}

\noindent
{\bf Acknowledgments}
This work was carried out on the Miracle Supercomputer, at the HiPerSPACE Computing Centre, UCL, which is funded by the U.K. Particle Physics and Astronomy Research
Council. We thank the anonymous referee for useful comments. BE aknowledges support from PPARC Grant PPA/G/S/1997/00728 and the award of a University of London Jubber Studentship. We thank Dr M. Rosa for making available to us a copy of the
photoionization code described by Och, Lucy and Rosa (1998).

\bibliographystyle{mn2e}
\bibliography{references}

\begin{thebibliography}{}

\bibitem[\protect\citeauthoryear{Abbott \& Lucy}{Abbott \&
  Lucy}{1985}]{abbott85}
Abbott D.~C.,  Lucy L.~B.,  1985, ApJ, 288, 679

\bibitem[\protect\citeauthoryear{Aggarwal}{Aggarwal}{1983}]{aggarwal83}
Aggarwal K.~M.,  1983, J.Phys.B., 16, 2405

\bibitem[\protect\citeauthoryear{Aggarwal}{Aggarwal}{1984a}]{aggarwal84a}
Aggarwal K.~M.,  1984a, SoPh, 94, 75

\bibitem[\protect\citeauthoryear{Aggarwal}{Aggarwal}{1984b}]{aggarwal84b}
Aggarwal K.~M.,  1984b, ApJS, 56, 303

\bibitem[\protect\citeauthoryear{Aldrovandi \& P\'equignot}{Aldrovandi \&
  P\'equignot}{1973}]{aldrovandi73}
Aldrovandi S.~M.,  P\'equignot D.,  1973, A\&A, 25, 137

\bibitem[\protect\citeauthoryear{Allen}{Allen}{1973}]{allen73}
Allen C.~W.,  1973, Astrophysical Quantities.
The Athlone Press, University of London, 4 Gower st., London

\bibitem[\protect\citeauthoryear{Almog \& Netzer}{Almog \&
  Netzer}{1989}]{almog89}
Almog Y.,  Netzer H.,  1989, MNRAS, 238, 57

\bibitem[\protect\citeauthoryear{Arnaud \& Raymond}{Arnaud \&
  Raymond}{1992}]{arnaud92}
Arnaud M.,  Raymond J.,  1992, ApJ, 398, 394

\bibitem[\protect\citeauthoryear{Baesgen, Diesch \& Grewing}{Baesgen
  et~al.}{1990}]{bassgen90}
Baesgen M.,  Diesch C.,    Grewing M.,  1990, 201, A\&A, 237

\bibitem[\protect\citeauthoryear{Balick, Hajian, Terzian, Perinotto \&
  Patriarchi}{Balick et~al.}{1998}]{balick98}
Balick B.,  Hajian A.~R.,  Terzian Y.,  Perinotto M.,    Patriarchi P.,  1998,
  AJ, 116, 360

\bibitem[\protect\citeauthoryear{Balick, Perinotto, Maccioni, Terzian \&
  Hajian}{Balick et~al.}{1994}]{balick94}
Balick B.,  Perinotto M.,  Maccioni A.,  Terzian Y.,    Hajian A.~R.,  1994,
  ApJ, 424, 800

\bibitem[\protect\citeauthoryear{Balick, Rugers, Terzian \& Chengalur}{Balick
  et~al.}{1993}]{balick93}
Balick B.,  Rugers M.,  Terzian Y.,    Chengalur J.~N.,  1993, ApJ, 411, 778

\bibitem[\protect\citeauthoryear{Balujia \& Zeippen}{Balujia \&
  Zeippen}{1988}]{baluja88}
Balujia K.~L.,  Zeippen C.~J.,  1988, J.Phys.B., 21, 1455

\bibitem[\protect\citeauthoryear{Bayes, Saraph \& Seaton}{Bayes
  et~al.}{1985}]{bayes85}
Bayes F.~A.,  Saraph H.~E.,    Seaton M.~J.,  1985, MNRAS, 215, 85

\bibitem[\protect\citeauthoryear{Benjamin, Skillman \& Smits}{Benjamin
  et~al.}{1999}]{benjamin99}
Benjamin R.~A.,  Skillman E.~D.,    Smits D.~P.,  1999, ApJ, 514, 307

\bibitem[\protect\citeauthoryear{Berrington}{Berrington}{1988}]{berrington88}
Berrington K.~A.,  1988, J.Phys.B., 21, 1083

\bibitem[\protect\citeauthoryear{Berrington, Burke, Dufton \&
  Kingston}{Berrington et~al.}{1981}]{berrington81}
Berrington K.~A.,  Burke P.~G.,  Dufton P.~L.,    Kingston A.~E.,  1981, ADNDT,
  26, 1

\bibitem[\protect\citeauthoryear{Bhatia, Feldman \& Doscheck}{Bhatia
  et~al.}{1979}]{bahtia79}
Bhatia A.~K.,  Feldman U.,    Doscheck G.~A.,  1979, A\&A, 80, 22

\bibitem[\protect\citeauthoryear{Bhatia \& Mason}{Bhatia \&
  Mason}{1980}]{bahtia80}
Bhatia A.~K.,  Mason H.~E.,  1980, MNRAS, 190, 925

\bibitem[\protect\citeauthoryear{Bjorkman \& Wood}{Bjorkman \&
  Wood}{2001}]{bjorkman01}
Bjorkman J.~E.,  Wood K.,  2001, ApJ, 554, 615

\bibitem[\protect\citeauthoryear{Boiss\'e}{Boiss\'e}{1990}]{boisse90}
Boiss\'e P.,  1990, A\&A, 228, 483

\bibitem[\protect\citeauthoryear{Brocklehurst}{Brocklehurst}{1972}]{brocklehur%
st72}
Brocklehurst M.,  1972, MNRAS, 157, 211

\bibitem[\protect\citeauthoryear{Brown \& Matthews}{Brown \&
  Matthews}{1970}]{brown70}
Brown R.~L.,  Matthews W.~G.,  1970, ApJ, 160, 939

\bibitem[\protect\citeauthoryear{Butler \& Zeippen}{Butler \&
  Zeippen}{1994}]{butler94}
Butler S.~E.,  Zeippen C.~J.,  1994, A\&AS, 108, 1

\bibitem[\protect\citeauthoryear{Castor}{Castor}{1974}]{castor74}
Castor J.~I.,  1974, ApJ, 189, 273

\bibitem[\protect\citeauthoryear{Corradi, Perinotto, Villaver, Mampaso \&
  Gon\c{c}alves}{Corradi et~al.}{1999}]{corradi99}
Corradi R. L.~M.,  Perinotto M.,  Villaver E.,  Mampaso A.,    Gon\c{c}alves
  D.~R.,  1999, ApJ, 523, 721

\bibitem[\protect\citeauthoryear{Drake \& Ulrich}{Drake \&
  Ulrich}{1980}]{drake80}
Drake G. W.~F.,  Ulrich H.,  1980, ApJS, 42, 351

\bibitem[\protect\citeauthoryear{Drake, Victor \& Dalgarno}{Drake
  et~al.}{1969}]{drake69}
Drake G. W.~F.,  Victor G.~A.,    Dalgarno A.,  1969, Phys.Rev., 180, 25

\bibitem[\protect\citeauthoryear{Dufton, Keenan, Hibbert, Stafford, Byrne \&
  Agnew}{Dufton et~al.}{1991}]{duftonetal91}
Dufton P.~L.,  Keenan F.~P.,  Hibbert A.,  Stafford R.~P.,  Byrne P.~B.,
  Agnew D.,  1991, MNRAS, 253, 474

\bibitem[\protect\citeauthoryear{Dufton \& Kingston}{Dufton \&
  Kingston}{1989}]{dufton89}
Dufton P.~L.,  Kingston A.~E.,  1989, MNRAS, 241, 209

\bibitem[\protect\citeauthoryear{Dufton \& Kingston}{Dufton \&
  Kingston}{1991}]{dufton91}
Dufton P.~L.,  Kingston A.~E.,  1991, MNRAS, 248, 827

\bibitem[\protect\citeauthoryear{Eidelsberg, Crifo-Magnant \&
  Zeippen}{Eidelsberg et~al.}{1981}]{eidelsberg81}
Eidelsberg M.,  Crifo-Magnant F.,    Zeippen C.~J.,  1981, A\&AS, 43, 455

\bibitem[\protect\citeauthoryear{Ercolano}{Ercolano}{2001}]{ercolano01}
Ercolano B.,  2001, Spectroscopic Challanges of Photoionized Plasmas, ASP Conf.
  Ser., 247, 281

\bibitem[\protect\citeauthoryear{Ercolano}{Ercolano}{2002}]{ercolano02}
Ercolano B.,  2002, Three Dimensional Monte Carlo Simulations of Ionized
  Nebulae, PhD Thesis, University of London

\bibitem[\protect\citeauthoryear{Ercolano, Morisset, Barlow, Storey \&
  Liu}{Ercolano et~al.}{2002}]{ercolanoII}
Ercolano B.,  Morisset C.,  Barlow M.~J.,  Storey P.~J.,    Liu X.-W.,  2002,
  MNRAS, submitted

\bibitem[\protect\citeauthoryear{Fang, Kwong \& Parkinson}{Fang
  et~al.}{1993}]{fang93}
Fang Z.,  Kwong V. H.~S.,    Parkinson W.~H.,  1993, ApJ, 413, 141

\bibitem[\protect\citeauthoryear{Ferland, Binette, Contini, Harrington,
  Kallman, Netzer, P\'equignot, Raymond, Rubin, Shields, Sutherland \&
  Viegas}{Ferland et~al.}{1995}]{ferland95}
Ferland G.,  Binette L.,  Contini M.,  Harrington J.,  Kallman T.,  Netzer H.,
  P\'equignot D.,  Raymond J.,  Rubin R.,  Shields G.,  Sutherland R.,
  Viegas S.,  1995, The Analysis of Emission Lines, STScI Symposium, 8, 83

\bibitem[\protect\citeauthoryear{Ferland}{Ferland}{1980}]{ferland80}
Ferland G.~J.,  1980, PASP, 92, 596

\bibitem[\protect\citeauthoryear{Fischer, Henning \& Yorke}{Fischer
  et~al.}{1994}]{fischer94}
Fischer O.,  Henning T.,    Yorke H.~W.,  1994, A\&A, 284, 187

\bibitem[\protect\citeauthoryear{Flannery, Roberge \& Rybicki}{Flannery
  et~al.}{1980}]{flannery80}
Flannery B.~P.,  Roberge W.,    Rybicki G.~B.,  1980, ApJ, 236, 598

\bibitem[\protect\citeauthoryear{Fleming, Bell, Hibbert, Vaeck \&
  Godefroid}{Fleming et~al.}{1996}]{fleming96}
Fleming J.,  Bell K.~L.,  Hibbert A.,  Vaeck N.,    Godefroid M.~R.,  1996,
  MNRAS, 279, 1289

\bibitem[\protect\citeauthoryear{Fleming, Brage, Bell, Vaeck, Hibbert,
  Godefroid \& Fischer}{Fleming et~al.}{1995}]{fleming95}
Fleming J.,  Brage T.,  Bell K.~L.,  Vaeck N.,  Hibbert A.,  Godefroid M.~R.,
   Fischer C.~F.,  1995, ApJ, 455, 758

\bibitem[\protect\citeauthoryear{Flower}{Flower}{1968}]{flower68}
Flower D.~R.,  1968, ApJL, 2, 205

\bibitem[\protect\citeauthoryear{Garc\'ia-Segura}{Garc\'ia-Segura}{1997}]{garc%
ia97}
Garc\'ia-Segura G.,  1997, ApJ, 489, 189

\bibitem[\protect\citeauthoryear{Gau \& Henry}{Gau \& Henry}{1977}]{gau77}
Gau J.~N.,  Henry R. J.~W.,  1977, PhRvA, 16, 986

\bibitem[\protect\citeauthoryear{Giles}{Giles}{1981}]{giles81}
Giles K.,  1981, MNRAS, 195, 63

\bibitem[\protect\citeauthoryear{Gruenwald, Viegas \& de Broguiere}{Gruenwald
  et~al.}{1997}]{gruenwald97}
Gruenwald R.,  Viegas S.~M.,    de Broguiere D.,  1997, ApJ, 480, 283

\bibitem[\protect\citeauthoryear{Harries \& Howarth}{Harries \&
  Howarth}{1997}]{harris97}
Harries T.~J.,  Howarth I.~D.,  1997, A\&AS, 121, 15

\bibitem[\protect\citeauthoryear{Harrington}{Harrington}{1968}]{harrington68}
Harrington J.~P.,  1968, ApJ, 152, 943

\bibitem[\protect\citeauthoryear{Hayes \& Nussbaumer}{Hayes \&
  Nussbaumer}{1984}]{hayes84}
Hayes M.~A.,  Nussbaumer H.,  1984, A\&A, 134, 193

\bibitem[\protect\citeauthoryear{Hummer}{Hummer}{1988}]{hummer88}
Hummer D.~G.,  1988, ApJ, 327, 477

\bibitem[\protect\citeauthoryear{Hummer}{Hummer}{1994}]{hummer94}
Hummer D.~G.,  1994, MNRAS, 268, 109

\bibitem[\protect\citeauthoryear{Hummer, Berrington, Eissner, Pradhan, Saraph
  \& Tully}{Hummer et~al.}{1993}]{hummer93}
Hummer D.~G.,  Berrington K.~A.,  Eissner W.,  Pradhan A.~K.,  Saraph H.~E.,
  Tully J.~A.,  1993, A\&A, 279, 298

\bibitem[\protect\citeauthoryear{Hummer \& Storey}{Hummer \&
  Storey}{1998}]{hummer98}
Hummer D.~G.,  Storey P.~J.,  1998, MNRAS, 297, 1073

\bibitem[\protect\citeauthoryear{Keenan, Feibelman \& Berrington}{Keenan
  et~al.}{1992}]{keenan92}
Keenan F.~P.,  Feibelman W.~A.,    Berrington K.~A.,  1992, ApJ, 389, 443

\bibitem[\protect\citeauthoryear{Kingdon \& Ferland}{Kingdon \&
  Ferland}{1996}]{kingdon96}
Kingdon J.~B.,  Ferland G.~J.,  1996, ApJS, 106, 205

\bibitem[\protect\citeauthoryear{Knigge, Woods \& Drew}{Knigge
  et~al.}{1995}]{knigge95}
Knigge C.,  Woods J.,    Drew J.,  1995, MNRAS, 273, 225

\bibitem[\protect\citeauthoryear{Landini \& {Monsignori Fossi}}{Landini \&
  {Monsignori Fossi}}{1990}]{landini90}
Landini M.,  {Monsignori Fossi} B.~C.,  1990, A\&AS, 82, 229

\bibitem[\protect\citeauthoryear{Landini \& {Monsignori Fossi}}{Landini \&
  {Monsignori Fossi}}{1991}]{landini91}
Landini M.,  {Monsignori Fossi} B.~C.,  1991, A\&AS, 91, 183

\bibitem[\protect\citeauthoryear{Lefevre, Bergeat \& Daniel}{Lefevre
  et~al.}{1982}]{lefevre82}
Lefevre J.,  Bergeat J.,    Daniel J.~V.,  1982, A\&A, 114, 341

\bibitem[\protect\citeauthoryear{Lefevre, Daniel \& Bergeat}{Lefevre
  et~al.}{1983}]{lefevre83}
Lefevre J.,  Daniel J.~V.,    Bergeat J.,  1983, A\&A, 121, 51

\bibitem[\protect\citeauthoryear{Lennon \& Burke}{Lennon \&
  Burke}{1991}]{lennon91}
Lennon D.~J.,  Burke V.~M.,  1991, MNRAS, 251, 628

\bibitem[\protect\citeauthoryear{Lopez, Meaburn \& Palmer}{Lopez
  et~al.}{1993}]{lopez93}
Lopez J.~A.,  Meaburn J.,    Palmer J.~W.,  1993, ApJ, 415, L135

\bibitem[\protect\citeauthoryear{Lucy}{Lucy}{1999}]{lucy99}
Lucy L.~B.,  1999, A\&A, 344, 282

\bibitem[\protect\citeauthoryear{Lucy}{Lucy}{2001}]{lucy01}
Lucy L.~B.,  2001, MNRAS, 326, 95L

\bibitem[\protect\citeauthoryear{Lucy}{Lucy}{2002}]{lucy02}
Lucy L.~B.,  2002, A\&A, 384, 725L

\bibitem[\protect\citeauthoryear{McLaughlin \& Bell}{McLaughlin \&
  Bell}{1998}]{mclaughlin98}
McLaughlin B.~M.,  Bell K.~L.,  1998, J.Phys.B., 31, 4317

\bibitem[\protect\citeauthoryear{Mendoza}{Mendoza}{1983}]{mendoza82}
Mendoza C.,  1983, Planetary Nebulae, IAU Symposium, ed. D.R. Flower, 103, 143

\bibitem[\protect\citeauthoryear{Mendoza \& Zeippen}{Mendoza \&
  Zeippen}{1983}]{mendoza83}
Mendoza C.,  Zeippen C.~J.,  1983, MNRAS, 202, 981

\bibitem[\protect\citeauthoryear{Monteiro, Morisset, Gruenwald \&
  Viegas}{Monteiro et~al.}{2000}]{monteiro00}
Monteiro H.,  Morisset C.,  Gruenwald R.,    Viegas S.~M.,  2000, ApJ, 537, 853

\bibitem[\protect\citeauthoryear{Morisset, Gruenwald \& Viegas}{Morisset
  et~al.}{2000}]{morisset00}
Morisset C.,  Gruenwald R.,    Viegas S.~M.,  2000, ApJ, 533, 931

\bibitem[\protect\citeauthoryear{Nahar}{Nahar}{2000}]{nahar00}
Nahar S.~N.,  2000, A\&AS, 147, 549

\bibitem[\protect\citeauthoryear{Nahar \& Pradhan}{Nahar \&
  Pradhan}{1999}]{nahar99}
Nahar S.~N.,  Pradhan A.~K.,  1999, A\&AS, 135, 347

\bibitem[\protect\citeauthoryear{Nussbaumer \& Rusca}{Nussbaumer \&
  Rusca}{1979}]{nussbaumer79}
Nussbaumer H.,  Rusca C.,  1979, A\&A, 72, 129

\bibitem[\protect\citeauthoryear{Nussbaumer \& Schmutz}{Nussbaumer \&
  Schmutz}{1984}]{nussbaumer84}
Nussbaumer H.,  Schmutz W.,  1984, A\&A, 138, 495

\bibitem[\protect\citeauthoryear{Nussbaumer \& Storey}{Nussbaumer \&
  Storey}{1981}]{nussbaumer81}
Nussbaumer H.,  Storey P.,  1981, A\&A, 96, 91

\bibitem[\protect\citeauthoryear{Nussbaumer \& Storey}{Nussbaumer \&
  Storey}{1982}]{nussbaumer82}
Nussbaumer H.,  Storey P.~J.,  1982, A\&A, 115, 205

\bibitem[\protect\citeauthoryear{Nussbaumer \& Storey}{Nussbaumer \&
  Storey}{1983}]{nussbaumer83}
Nussbaumer H.,  Storey P.~J.,  1983, A\&A, 126, 75

\bibitem[\protect\citeauthoryear{Nussbaumer \& Storey}{Nussbaumer \&
  Storey}{1986}]{nussbaumer86}
Nussbaumer H.,  Storey P.~J.,  1986, A\&AS, 64, 545

\bibitem[\protect\citeauthoryear{Nussbaumer \& Storey}{Nussbaumer \&
  Storey}{1987}]{nussbaumer87}
Nussbaumer H.,  Storey P.~J.,  1987, A\&AS, 69, 123

\bibitem[\protect\citeauthoryear{Och, Lucy \& Rosa}{Och et~al.}{1998}]{och98}
Och S.~R.,  Lucy L.~B.,    Rosa M.~B.,  1998, A\&A, 336, 301

\bibitem[\protect\citeauthoryear{Osterbrock}{Osterbrock}{1989}]{osterbrock89}
Osterbrock D.~E.,  1989, Astrophysics of Gaseous Nebulae and Active Galactic
  Nuclei.
University Science Books, Mill Valley, CA

\bibitem[\protect\citeauthoryear{Osterbrock \& Wallace}{Osterbrock \&
  Wallace}{1977}]{osterbrock77}
Osterbrock D.~E.,  Wallace R.~K.,  1977, ApL, 19, 110

\bibitem[\protect\citeauthoryear{Pengelly}{Pengelly}{1964}]{pengelly64}
Pengelly R.~M.,  1964, MNRAS, 127, 145

\bibitem[\protect\citeauthoryear{P\'equignot}{P\'equignot}{1986}]{pequignot86}
P\'equignot D.,  1986, in Workshop on model nebulae Publication de
  l'Observatoire de Meudon, Paris, ed. D. Pequignot

\bibitem[\protect\citeauthoryear{P\'equignot \& Aldrovandi}{P\'equignot \&
  Aldrovandi}{1976}]{pequignot76}
P\'equignot D.,  Aldrovandi S.~M.,  1976, A\&A, 50, 141

\bibitem[\protect\citeauthoryear{P\'equignot, Ferland, Netzer, Kallman,
  Ballantyne, Dumont, Ercolano, Harrington, Kraemer, Morisset, Nayakshin, Rubin
  \& Sutherland}{P\'equignot et~al.}{2001}]{pequignot01}
P\'equignot D.,  Ferland G.,  Netzer H.,  Kallman T.,  Ballantyne D.,  Dumont
  A.-M.,  Ercolano B.,  Harrington P.,  Kraemer S.,  Morisset C.,  Nayakshin
  S.,  Rubin R.,    Sutherland R.,  2001, Spectroscopic Challanges of
  Photoionized Plasmas, ASP Conference Series, 247, 533

\bibitem[\protect\citeauthoryear{P\'equignot, Petitjean \& Boisson}{P\'equignot
  et~al.}{1991}]{pequignot91}
P\'equignot D.,  Petitjean P.,    Boisson C.,  1991, A\&A, 251, 680

\bibitem[\protect\citeauthoryear{P\'equignot, Stasinska \&
  Aldrovandi}{P\'equignot et~al.}{1978}]{pequignot78}
P\'equignot D.,  Stasinska G.,    Aldrovandi S. M.~V.,  1978, A\&A, 63, 313

\bibitem[\protect\citeauthoryear{Perinotto}{Perinotto}{2000}]{perinotto00}
Perinotto M.,  2000, ApSS, 274, 205

\bibitem[\protect\citeauthoryear{Pradhan}{Pradhan}{1976}]{pradhan76}
Pradhan A.~K.,  1976, MNRAS, 177, 31

\bibitem[\protect\citeauthoryear{Ramsbottom, Bell \& Keenan}{Ramsbottom
  et~al.}{1998}]{ramsbottom98}
Ramsbottom C.~A.,  Bell K.~L.,    Keenan F.~P.,  1998, MNRAS, 293, 233

\bibitem[\protect\citeauthoryear{Reilman \& Manson}{Reilman \&
  Manson}{1979}]{reilman79}
Reilman R.~F.,  Manson S.~T.,  1979, ApJS, 40, 815

\bibitem[\protect\citeauthoryear{Robbins}{Robbins}{1968}]{robbins68}
Robbins R.~R.,  1968, ApJ, 151, 497

\bibitem[\protect\citeauthoryear{Rubin}{Rubin}{1968}]{rubin68}
Rubin R.~H.,  1968, ApJ, 153, 761

\bibitem[\protect\citeauthoryear{Sahai}{Sahai}{2000}]{sahai00}
Sahai R.,  2000, ApJ, 537, L43

\bibitem[\protect\citeauthoryear{Saraph \& Storey}{Saraph \&
  Storey}{1996}]{saraph96}
Saraph H.~E.,  Storey P.~J.,  1996, A\&AS, 115, 151

\bibitem[\protect\citeauthoryear{Shull \& van Steenberg}{Shull \& van
  Steenberg}{1982}]{shull82b}
Shull J.~M.,  van Steenberg M.,  1982, ApJS, 49, 351

\bibitem[\protect\citeauthoryear{Soker}{Soker}{1997}]{soker97}
Soker N.,  1997, ApJS, 112, 487

\bibitem[\protect\citeauthoryear{Soker}{Soker}{2001}]{soker01}
Soker N.,  2001, MNRAS, 324, 699

\bibitem[\protect\citeauthoryear{Stafford, Bell, Hibbert \&
  Wijesundera}{Stafford et~al.}{1994}]{stafford94}
Stafford R.~P.,  Bell K.~L.,  Hibbert A.,    Wijesundera W.~P.,  1994, MNRAS,
  268, 816

\bibitem[\protect\citeauthoryear{Storey}{Storey}{1981}]{storey81}
Storey P.~J.,  1981, MNRAS, 195, 27

\bibitem[\protect\citeauthoryear{Storey \& Hummer}{Storey \&
  Hummer}{1995}]{storey95}
Storey P.~J.,  Hummer D.~G.,  1995, MNRAS, 272, 41

\bibitem[\protect\citeauthoryear{Thomas \& Nesbit}{Thomas \&
  Nesbit}{1975}]{thomas75}
Thomas Nesbit 1975, PhRvA, 12, 2378

\bibitem[\protect\citeauthoryear{Verner, Ferland, Korista \& Yakovlev}{Verner
  et~al.}{1996}]{verner96}
Verner D.~A.,  Ferland G.~J.,  Korista K.~T.,    Yakovlev D.~G.,  1996, ApJ,
  465, 487

\bibitem[\protect\citeauthoryear{Verner \& Yakovlev}{Verner \&
  Yakovlev}{1995}]{verner95}
Verner D.~A.,  Yakovlev D.~G.,  1995, A\&AS, 109, 125

\bibitem[\protect\citeauthoryear{Wiese, Smith \& Glennon}{Wiese
  et~al.}{1966}]{wiese66}
Wiese W.~L.,  Smith M.~W.,    Glennon B.~M.,  1966, Atomic transition
  probabilities: Hydrogen through Neon. A critical data compilation.
NSRDS-NBS 4, Washington, D.C.: US Department of Commerce, National Buereau of
  Standards

\bibitem[\protect\citeauthoryear{Wolf, Henning \& Secklum}{Wolf
  et~al.}{1999}]{wolf99}
Wolf S.,  Henning T.,    Secklum B.,  1999, A\&A, 349, 839

\bibitem[\protect\citeauthoryear{Zeippen}{Zeippen}{1982}]{zeippen82}
Zeippen C.~J.,  1982, MNRAS, 198, 111

\bibitem[\protect\citeauthoryear{Zhang, Graziani \& Pradhan}{Zhang
  et~al.}{1994}]{zhang94}
Zhang H.~L.,  Graziani M.,    Pradhan A.~K.,  1994, A\&A, 283, 319

\end{thebibliography}

\vspace{7mm}

\noindent
{\bf Appendix A: Atomic Data References} \\
{\noindent} Free-bound emission for hydrogenic ions (H~{\sc i} and He~{\sc ii}): 
\citet{ferland80}

{\noindent} Free-bound emission for He~{\sc i}:
\citet{brown70}

{\noindent} Two-photon emission for hydrogenic ions (H~{\sc i} and He~{\sc ii}):
\citet{nussbaumer84}

{\noindent} Two-photon emission for He~{\sc i}:
\citet{drake69}

{\noindent} Free-free emission for interaction between ions of nucleus Z and electrons: 
\citet{allen73}

{\noindent} Effective recombination coefficient to H~{\sc i}~2$^2$S:
\citet{pengelly64}

{\noindent} Effective recombination coefficient to He~{\sc i}~2$^1$S:
\citet{almog89}

{\noindent} H~{\sc i} and He~{\sc ii} recombination line emissivities:
\citet{storey95}

{\noindent} He~{\sc i} recombination line emissivities:
\citet{benjamin99}

{\noindent} Collision transition rates for H~{\sc i} 2$^2$S~-~2$^2$P:
\citet[][page94]{osterbrock89}

{\noindent} Cooling due to free-free radiation from hydrogenic ions (H~{\sc i} and He~{\sc ii}): 
\citet{hummer94}

{\noindent} Cooling due to free-free radiation from He~{\sc i}: 
\citet{hummer98}

{\noindent} Cooling due to recombination of hydrogenic ions (H~{\sc i} and He~{\sc ii}): 
\citet{hummer94}

{\noindent} Cooling due to He~{\sc i} recombination: 
\citet{hummer98}

{\noindent} Collision Ionization of hydrogen: 
\citet{drake80}

{\noindent} Charge exchange with hydrogen:
\citet{kingdon96}

{\noindent} Fits to calculate rates of radiative recombination for H-like, He-like, Li-like, Ne-like ions:
\citet{verner96}. Other ions of C, N, O, Ne:
\citet{pequignot91}.
Fe~{\sc xvii}-{\sc xxiii}:
\citet{arnaud92}.
Other ions of Mg, Si, S, Ar, Ca, Fe, Ni:
\citet{shull82b}.
Other ions of Na, Al: 
\citet{landini90}.
Other ions of F, P, Cl, K, Ti, Cr, Mn, Co (excluding Ti~{\sc I}-{\sc ii} and Cr~{\sc I}-{\sc iv}):
\citet{landini91}

{\noindent} Dielectronic recombination coefficients:
\citet{nussbaumer83, nussbaumer86, nussbaumer87} 

{\noindent} Non relativistic free-free Gaunt factor for hydrogenic ions:
\citet{hummer88}

{\noindent} Fits to opacity project data for the photoionization cross-sections (outer shell):
\citet{verner96}

{\noindent} Collision strengths, and transition probabilities to calculate collisionally excited line strengths from ions:
\begin{description}
  \item[C~{\sc i}] Collision strengths from \citet{pequignot76}; $^5$S-$^3$P from \citet{thomas75}. Transition probabilities from \citet{nussbaumer79}.
   \item[C~{\sc ii}] Collision strengths from \citet{hayes84}. Transition probabilities from \citet{nussbaumer81}.
   \item[C~{\sc iii}] collision strengths and transition probabilities from \citet{keenan92} and \citet{fleming96}.
   \item[C~{\sc iv}] collision strengths from \citet{gau77}. Transition probabilities from \citet{wiese66}. 
   \item[Mg~{\sc i}] Collision Strengths from Saraph (1986) JAJOM calculations. Transition probabilities from \citet{mendoza82}.
   \item[Mg~{\sc ii}] collision strengths and transition probabilities from \citet{mendoza82}.
   \item[Mg~{\sc iv}] Collision strengths from \citet{butler94}. Transition probabilities from \citet{mendoza83}.
   \item[Mg~{\sc v}]  Collision strengths from \citet{butler94}. Transition probabilities from \citet{mendoza82}.
   \item[Mg~{\sc vi}] Collision strengths from  \citet{bahtia80}. Transition probabilities from \citet{eidelsberg81}.
   \item[Mg~{\sc vii}] Collision strengths from \citet{aggarwal84a} and \citet{aggarwal84b}. Transition probabilities from \citet{nussbaumer79}.
   \item[Ne~{\sc ii}] Collision strength from \citet{bayes85}. Transition probabilities from \citet{mendoza83}.
   \item[Ne~{\sc iii}] Collision strengths from \citet{butler94}. Transition probabilities from \citet{mendoza83}.
   \item[Ne~{\sc iv}] Collision strengths from \citet{giles81}. Transition probabilities from \citet{zeippen82}.
   \item[Ne~{\sc v}] Collision strengths from \citet{lennon91}. Transition probabilities from \citet{nussbaumer79}. 
   \item[Ne~{\sc vi}] Collision strengths from Butler \& Storey (unpublished). Transition probabilities from \citet{wiese66}. 
   \item[N~{\sc i}] Collision strengths from \citet{berrington81}. Transition probabilities from \citet{zeippen82}. 
   \item[N~{\sc ii}] Collision strengths from \citet{stafford94}. Transition probabilities from \citet{nussbaumer79}.
   \item[N~{\sc iii}] Collision strengths from \citet{nussbaumer79}, rescaled to \citet{nussbaumer82}, fine-structure terms from Butler \& Storey (unpublished). Transition probabilities from \citet{fang93}. 
   \item[N~{\sc iv}]  Collision strengths from \citet{mendoza83}. Transition probabilities from \citet{nussbaumer79} and \citet{fleming95}.
   \item[N~{\sc v}] Collision strengths  from \citet{osterbrock77}. Transition probabilities from \citet{wiese66}.
   \item[O~{\sc i}] Collision strengths from \citet{berrington81} and \citet{berrington88}. Transition probabilities from \citet{baluja88}.
   \item[O~{\sc ii}] Collision strengths from \citet{pradhan76}. Transition probabilities from \citet{zeippen82}.
   \item[O~{\sc iii}] Collision strengths from \citet{aggarwal83}. Transition probabilities from \citet{nussbaumer81}.
   \item[O~{\sc iv}] Collision strengths from \citet{zhang94} and from \citet{hayes84}. Transition probabilities from \citet{nussbaumer82}.
   \item[O~{\sc v}] Collision strengths and transition probabilities from \citet{mendoza82}.
   \item[O~{\sc vi}] Collision strengths and transition probabilities from \citet{mendoza82}.
   \item[Si~{\sc ii}] Collision strengths from \citet{dufton91}. Transition probabilities from \citet{mendoza83} and from \citet{duftonetal91}. 
   \item[Si~{\sc iii}] Collision strengths from \citet{dufton89}. Transition probabilities from \citet{mendoza83} 
   \item[Si~{\sc iv}] Collision strengths and transition probabilities from \citet{mendoza82}.
   \item[Si~{\sc vii}] Fine structure collision strengths from Butler (unpublished). Transition probabilities from \citet{bahtia79}.
   \item[S~{\sc ii}] Collision strengths from \citet{mendoza83}. Transition probabilities from \citet{mendoza82}. 
   \item[S~{\sc iii}] Collision strengths from \citet{mendoza83}. Transition probabilities from \citet{mendoza82} 
   \item[S~{\sc iv}] Collision strengths from \citet{saraph96}. Transition probabilities from Storey (unpublished)

\end{description}

\end{document}